\documentclass[final,3p,times]{elsarticle}

\usepackage{graphicx}
\usepackage{hyperref}
\usepackage{afterpage}
\usepackage{miller}
\usepackage{amssymb}
\usepackage{amsmath}
\usepackage{footnote}

\journal{European Journal of Mechanics - A/Solids}

\begin{document}

\begin{frontmatter}



\title{Elastic properties of solid material with various arrangements of spherical voids}

 \author[label1]{Sascha Heitkam}
 \author[label2]{Wiebke Drenckhan}
 \author[label3]{Thomas Titscher}
 \author[label4]{Denis Weaire}
 \author[label5]{Daniel Christopher Kreuter}
 \author[label6]{David Hajnal}
 \author[label2]{Frederic Piechon}
 \author[label1]{Jochen Fr\"ohlich}
 \address[label1]{Institute of Fluid Mechanics, Technische Universit\"at Dresden, 01069 Dresden, Germany. }
 \address[label2]{Laboratoire de Physique des Solides, CNRS, Université Paris-Sud, 91405 Orsay, France. }
 \address[label3]{BAM Federal Institute for Materials Research and Testing, 12205 Berlin, Germany. }
 \address[label4]{School of Physics, Trinity College Dublin 2, Ireland. }
 \address[label5]{Institut f\"ur Festk\"orpermechanik, Technische Universit\"at Dresden, 01307 Dresden, Germany. }
 \address[label6]{BASF SE, 67056 Ludwigshafen, Germany. }

%\author{Thomas Titscher, Sascha Heitkam, Daniel Kreuter, Wiebke Drenckhan, David Hajnal, Frederic Piechon, Jochen Fr\"ohlich}

%\address{adress}

\begin{abstract}

In this work the linear elastic properties of materials containing spherical voids are calculated and compared using finite element simulations. The focus is on homogeneous solid materials with spherical, empty voids of equal size. The voids are arranged on crystalline lattices (SC, BCC, FCC and HCP structure) or randomly, and may overlap in order to produce connected voids. In that way, the entire range of void fraction between 0.00 and 0.95 is covered, including closed-cell and open-cell structures. For each arrangement of voids and for different void fractions the full stiffness tensor is computed. From this, the Young's modulus and Poisson ratios are derived for different orientations. Special care is taken of assessing and reducing the numerical uncertainty of the method. In that way, a reliable quantitative comparison of different void structures is carried out. Among other things, this work shows that the Young's modulus of FCC in the \hkl (1 1 1) plane differs from HCP in the \hkl (0 0 0 1) plane, even though these structures are very similar. For a given void fraction SC offers the highest and the lowest Young's modulus depending on the direction. For BCC at a critical void fraction a switch of the elastic behaviour is found, as regards the direction in which the Young's modulus is maximised. For certain crystalline void arrangements and certain directions Poisson ratios between 0 and 1 were found, including values that exceed the bounds for isotropic materials. For subsequent investigations the full stiffness tensor for a range of void arrangements and void fractions are provided in the supplemental material.

\end{abstract}


\end{frontmatter}

\tableofcontents

\section{Introduction}
\label{sec:introduction}

Introducing spherical voids into a solid and otherwise homogeneous and isotropic material changes its mechanical properties significantly. Such materials are currently generated using templating techniques~\cite{Yin2012}, by integrating hollow spheres in a matrix~\cite{Kenig1985} or by direct foaming~\cite{Testouri2010, Testouri2012}. The voids of the resulting material can be arranged in an ordered or in a disordered manner. It is therefore of considerable interest to compare the mechanical properties of different ordered and disordered void structures, including the case where the voids overlap.\\
\begin{figure}[htbp]
	\centering
		\includegraphics[width=.7\textwidth]{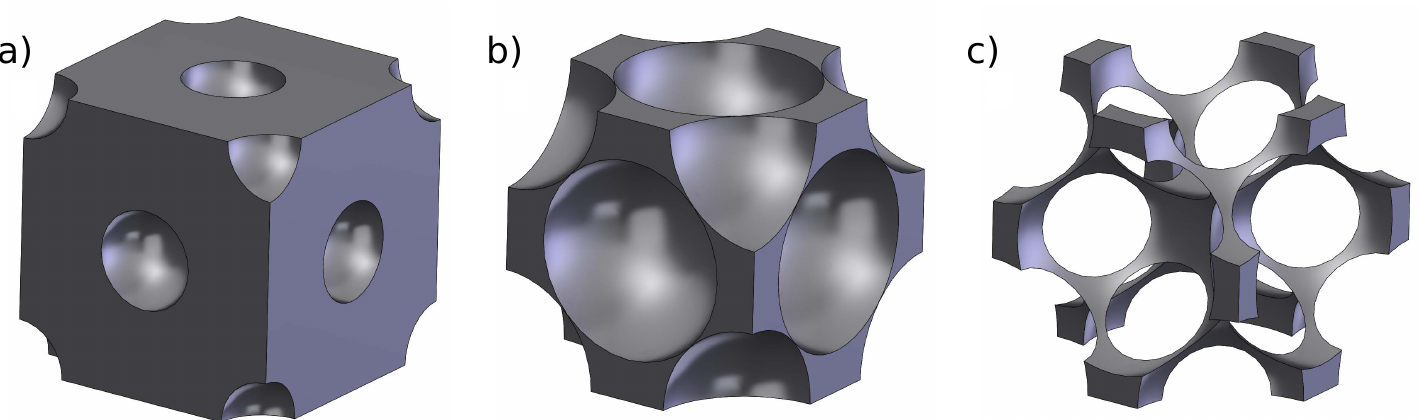}
	\caption{Void material with spherical voids of equal size in face-centred cubic arrangement at different void fractions. a) closed-cell structure at $\phi_v = 0.09$, b) touching spheres at $\phi_v = 0.74$ and c) open-cell structure at $\phi_v = 0.95$.  }
	\label{fig:intro}
\end{figure}
When direct foaming is used, the equal-volume bubbles tend to organize themselves on a close-packed lattice in densest packing~\cite{Heitkam2012}. Depending on the precise method of generation, the void arrangement may be chosen to be dominated by FCC (face-centred cubic) or HCP (hexagonal close-packed) arrangement~\cite{Heitkam2012, Drenckhan2010}. This raises the practically relevant question whether one of these arrangements should be preferred over the other, due to advantageous mechanical properties of the resulting solid void material.\\
Porous materials show a very rich range of non-linear mechanical behaviour, including plastic deformation, buckling and rupture. Here, we concentrate on the linear elastic behaviour, corresponding to infinitesimally small strain. The complementary problem, the elastic properties of crystalline arrangements of solid spheres, has been investigated experimentally and numerically by several authors~\cite{Radjai1999, OHern2001, Mueggenburg2002, Sanders2003, Ngan2004, Yin2012, An2013}. The elastic properties of the void material of these packings, however, have not yet been investigated sufficiently and comparatively.\\
Before computers made their breakthrough in science, the elastic properties of void material were estimated by superposition of the effects of a single void~\cite{Hill1965, Budiansky1965, Iwakuma1983}. These methods yield good results for low void fraction. However, with increasing void fraction, higher orders of interaction between the voids have to be taken into account~\cite{Eischen1993, Torquato1997, Torquato1998}. Christensen~\cite{Christensen1990} compared different micro-mechanic models available at that time.
In 1992, Day et al.~\cite{Day1992} developed a simple Finite Element Method (FEM) to calculate the elastic properties of a two-dimensional material with circular voids. They investigated the influence of void fraction and topology separately and devised a simple analytical explanation for the calculated values. After 1992, increasing computer power became available for many research groups, resulting in further direct numerical simulations of the interstitial material of sphere or bubble arrangements in three dimensions~\cite{Segurado2002, Ni2007, Bouhlel2010, Saadatfar2012}. In 2006, Ni et al.~\cite{Ni2007} calculated the Young's modulus of a simple cubic void structure and compared their results to analytical estimations of~\cite{Iwakuma1983} and~\cite{Cohen2004}, which are used later for comparison.\\
The agreement between analytical~\cite{Cohen2004, Iwakuma1983} and numerical~\cite{Ni2007} methods was very good. However, the graphs of Young's modulus versus void fraction depend only weakly on the structure. Thus, small derivations between the graphs raise the question, as to whether a difference results from the uncertainty of the method or rather from the structural differences of the investigated materials. In order to reliably extract the structural effects, one therefore needs to apply an identical numerical method to different structures, taking great care of the numerical uncertainty. Additionally, many of the available studies are confined to low or medium void fractions. \\
In this paper, a comparative study of a wide variety of dense sphere packings is carried out, revealing the influence of the structure on the elastic properties. The entire range of void fractions is considered, as illustrated in Figure~\ref{fig:intro}. Small voids form closed-cell void material with low void fraction. Retaining the regularly arranged void centres and increasing the diameter the voids touch each other at a certain void fraction $\phi_{tv,ouch}$, forming closed-packed void material. At even higher void fractions, the voids overlap, forming open-cell void materials.

\section{Material and Methods}
\label{sec:material_and_methods}
\subsection{Definition of sphere structures}
\begin{figure*}[htbp]
	\centering
		\includegraphics[width=0.9\textwidth]{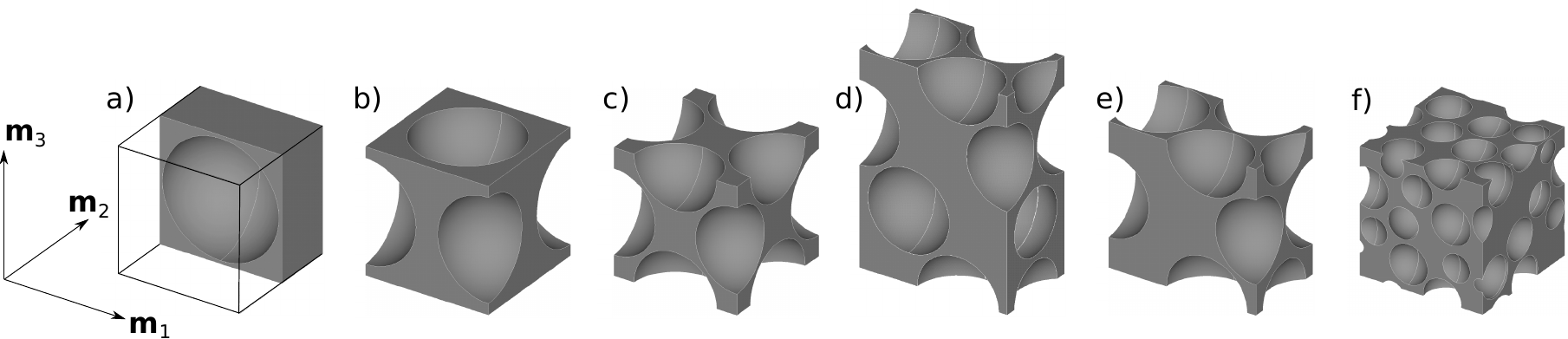}
	\caption{Sketch of the chosen RVEs and their orientation in the original basis $\left\{ \mathbf{m}_1, \mathbf{m}_2, \mathbf{m}_3 \right\}$, from left to right: a) Simple cubic (SC), b) Body-centred cubic (BCC), c) Face-centred cubic (FCC), d) Hexagonal face-centred cubic (FCCh), e) Hexagonal close-packed (HCP), f) Random close-packed (RCP). Corresponding parameters are given in Table~\ref{tab:RVEs}}
	\label{fig:RVE}
\end{figure*}

\begin{figure}[htbp]
	\centering
		\includegraphics[width=.9\textwidth]{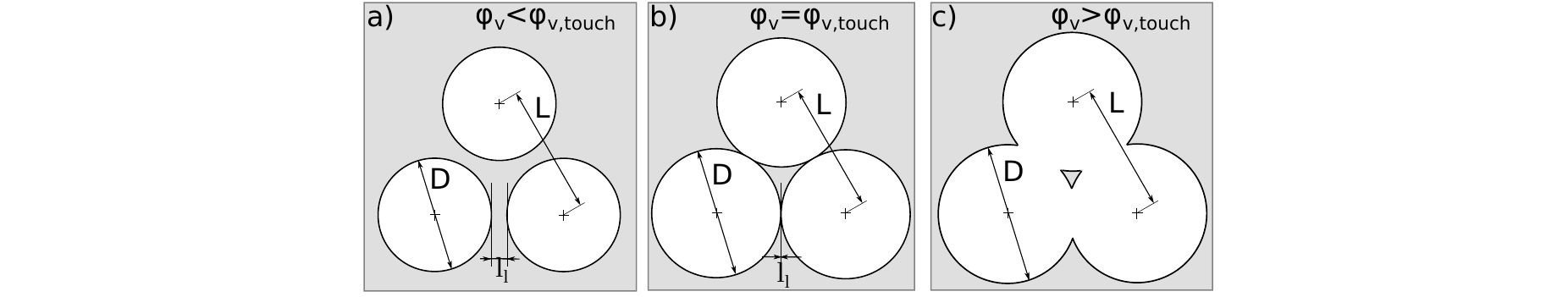}
	\caption{Basic geometry of sphere packing, showing lattice spacing $L$, Void diameter $D$, and void separation $l_l$. Depending on the void diameter the structure is a) closed-cell, b) indefinite or c) open-cell. }
	\label{fig:fig2}
\end{figure}

Monodisperse spheres or microbubbles tend to crystallise when they become agglomerated. This means that their centres form a periodic, crystalline lattice. Since these systems strive for densest packing, they are usually arranged in the hexagonally close-packed (HCP) or face-centred cubic (FCC) structure, both providing equally dense sphere packings~\cite{Weaire2008}. For comparison, simple cubic (SC) and body centred cubic (BCC) arrangements are also taken into account here. If the spheres are slightly polydisperse or if the agglomeration process is too fast to allow for relaxation, random closed-packed (RCP) structures are created.\\
From the different structures mentioned above, rectangular or cubic representative volume elements (RVE) were derived which are shown in Figure~\ref{fig:RVE}. Except for SC, the RVE does not coincide with the primitive cell of the crystalline arrangements. Rather, it is the smallest cuboid cell which may be periodically combined to represent the complete structure, because the numerical method only allows for orthogonal, periodic boundaries. Parameters of the chosen RVE are given in Table~\ref{tab:RVEs}. Note that for FCC two different RVEs were applied and compared. The cubic RVE, labelled FCC, is a cube, bounded by planes in $(100)$, $(010)$ and $(001)$. The hexagonal RVE, labelled FCCh, is a cuboid, bounded by $(111)$, $(1\bar{1}0)$ and  $(11\bar{2})$ planes. This provides an additional test of the method applied by comparing the Young's moduli of the different RVE of the same structure. This is explained in more detail in Section~\ref{sec:validation} below. 
The RCP structure is special, since it does not correspond to a crystalline lattice, but it does involve periodic boundary conditions. The sphere positions for this case were generated using a gas-dynamic algorithm that is freely available~\cite{Skoge2006}. Drugan et al.~\cite{Drugan1996, Drugan2000} found, that with six spheres in an RVE of disordered voids, the statistical uncertainty of the mechanical properties is below 5\%. Aiming for very high accuracy, here RVEs with 30 spheres were generated. The statistical uncertainty resulting from this type of RVE was investigated, as reported in Section~\ref{sec:validation} below.\\ 
\begin{table*}[htbp]
\begin{tabular}{cccccc}
structure 						 & label &$(L_x \times L_y \times L_z)/L$                   & $N_v$      & $\phi_{v,\mathrm{touch}}$  & $\phi_{v,\max}$ \\ \hline
simple cubic			 		 & SC	   &$1 \times 1 \times 1$			                        &	$1$			 &  $\frac{1}{6}\pi    \approx 52{,}4\% $ & $\approx 96{,}5\% $   \\
body centred cubic	   & BCC	 &$\frac{2}{\sqrt{3}} \times \frac{2}{\sqrt{3}} \times \frac{2}{\sqrt{3}}$	                                                                                                                     & 	$2$			&  $\frac{\sqrt{3}}{8}\pi \approx 68\% $&  $\approx 99{,}5\% $ \\
face centred cubic 		 & FCC	 &$\sqrt{2} \times \sqrt{2} \times \sqrt{2}$	      &	$4$			 &  $\frac{\sqrt{2}}{6}\pi \approx 74\% $ & $\approx 99{,}4\% $ \\
hexagonal FCC 				 & FCCh  &$1 \times \sqrt{3} \times 3 \sqrt{\frac{2}{3}}$	  &  $6$		 &  $\frac{\sqrt{2}}{6}\pi \approx 74\% $ & $\approx 99{,}4\% $  \\
hexagonal close-packed & HCP   &$	1 \times \sqrt{3} \times 2 \sqrt{\frac{2}{3}}$  & 	$4$	   &  $\frac{\sqrt{2}}{6}\pi \approx 74\% $ & $\approx 99{,}4\% $ \\	
random                 & RCP   &$\approx 3.1 \times 3.1 \times 3.1$               & $30$     &  $ \approx 62 \%$& N.A.  %\\
%PRIME & PRIME & $5 \times 10 \times 5 $ & $250$ & $250$ & $a$ & $ \frac{1}{6}\pi \approx 52{,}4\%  $ 
\end{tabular}
\caption{Parameters of the RVEs of the structures considered. The size of the RVE in $x$-, $y$- and $z$- direction is denoted $L_x$, $L_y$, and $L_z$, respectively, while $N_v$ is the number of spheres in each RVE, $\phi_{v,\mathrm{touch}}$ the void fraction for touching spheres and $\phi_{v,\max}$ the void fraction for disintegration of the material.}
\label{tab:RVEs}
\end{table*}
The solid fraction $\phi_s = V_{solid}/V_{RVE}$ of a void material is the ratio of the volume of solid material $V_{solid}$ contained in a given RVE with the volume $V_{RVE}$. The void fraction $\phi_v = 1 - \phi_s$ is the ratio between the void volume $V_{void}$ and the total volume of the RVE. In case of separated spherical voids, the void volume can be calculated from the sum of the volume of each spherical void contained in a given RVE. In this case, the void fraction depends on the sphere diameter $D$, the lattice spacing $L$, and the packing density $\phi_{v,\mathrm{touch}}$ for touching spheres of the structure considered
\begin{equation}
1-\phi_s = \phi_v = \frac{V_{void}}{V_{RVE}} = \phi_{v,\mathrm{touch}}  \left(\frac{D}{L} \right)^3 = \phi_{v,\mathrm{touch}}  \left(1 - \frac{l_l}{L} \right)^3 .
\label{eq:phi}
\end{equation}
Defining the separation of two voids to be $l_l = L - D$ as displayed in Figure~\ref{fig:fig2} yields the last equality in Equation~(\ref{eq:phi}).
For void fractions above $\phi_{v,\mathrm{touch}}$ the spherical voids overlap, yielding negative values for $l_l$. Due to that, Equation~(\ref{eq:phi}) is not valid for overlapping voids.

\subsection{Computation of elastic properties}

The goal of the method elaborated in this study is to find an equivalent homogeneous continuum material that is, in a volume-averaged sense, elastically equivalent to the heterogeneous RVE. The method of homogenization is well understood and documented, for example in~\cite{Nemat-Nasser2013}. The linear elastic behaviour is described by Hooke's law, which can be written as
\begin{equation}
\mathbf{\sigma} = \mathbf{C}  \mathbf{\varepsilon} .%\left< \mathbf{\sigma} \right>=\mathbf{C} \left< \mathbf{\varepsilon} \right> ,
\label{eq:hooke}
\end{equation}
This law relatess the applied stress $\mathbf{\sigma}$, the resulting strain $\mathbf{\varepsilon}$ and the stiffness tensor $\mathbf{C}$. For the equivalent homogeneous continuum material it is required that it stores the same elastic energy $U$ per Volume $V$ as the RVE when applying the same global strain, i.e. 
\begin{equation}
\frac{U}{V}= \frac{1}{2}  \mathbf{\varepsilon}^T \mathbf{C}  \mathbf{\varepsilon} .
\label{eq:energy}
\end{equation}
Stress, strain and the stiffness tensor are expressed in Voigt's notation in the following. The strain vector $\mathbf{\varepsilon}$ consists of 6 elements, which are normal strain $\varepsilon_{11}$, $\varepsilon_{22}$, $\varepsilon_{33}$ and shear strain $\varepsilon_{12}$, $\varepsilon_{23}$ and $\varepsilon_{31}$. The stress vector $\mathbf{\sigma}$ contains the corresponding 6 elements. Due to symmetry, the 6$\times$6 elements of $\mathbf{C}$ consist of 21 independent elements. Starting from a heterogeneous RVE, these can be computed by applying 21 independent load cases and calculating the corresponding elastic energy $U_{ij}$. These load cases are defined by 21 independent sets of strain $\mathbf{\varepsilon}$. One has to apply 6 sets of strain with one non-zero element $\varepsilon_k \neq 0$ yielding $U_{kk}$ and 15 sets of strain with two non-zero elements $\varepsilon_k \neq 0$, $\varepsilon_l \neq 0$ yielding $U_{kl}$. Note, that $U_{kl}$ is not a tensor but the scalar value of the elastic energy corresponding to the load case $\varepsilon_k \neq 0$, $\varepsilon_l \neq 0$.   \\
The diagonal elements $C_{kk}$ can be determined from sets of strain with only one element $\varepsilon_k \neq 0$
\begin{equation}
C_{kk} = 2\frac{U_{kk}}{\varepsilon_k^2 \; V} .
\label{eq:Ckk}
\end{equation}
The off-diagonal elements $C_{kl}$ result from sets with two strains $\varepsilon_k \neq 0$, $\varepsilon_l \neq 0$
\begin{equation}
C_{kl}=\frac{U_{kl} - U_{kk} - U_{ll} }{\varepsilon_k \; \varepsilon_l \; V} .
\label{eq:Ckl}
\end{equation}
Note, that $C_{kl} = - C_{lk}$ holds.\\
From the stiffness tensor $C_{kl}$ one obtains the compliance tensor $\mathbf{D}$ = $\mathbf{C}^{-1}$ by calculating its inverse. One obtains a Young's modulus $E_{ii}$ of the RVE from the first three main diagonal elements of the compliance tensor via
\begin{equation}
E_{11}=\frac{1}{D_{11}} \text{,} \quad E_{22}=\frac{1}{D_{22}}\text{,} \quad E_{33}=\frac{1}{D_{33}}.
\label{eq:Young's}
\end{equation}
These refer to the chosen basis of the compliance tensor but they can be rotated to obtain a value for any direction, as described below.
The Young's modulus of the void structure, normalized by the Young's modulus of the matrix material $E_0$, is one main material parameter considered in the present study. Voigt's rule of mixture~\cite{Voigt1889} provides an upper bound for the Young's modulus of a porous material
\begin{equation}
\frac{E_{ii}}{E_0} \leq (1-\phi_v) = \phi_s .
\label{eq:voigt}
\end{equation}
Poisson ratios $\nu_{ij}$ of the RVEs can be extracted from the compliance tensor 
\begin{equation}
 \nu_{ij} = - \frac{D_{ji}}{D_{jj}}.
\label{eq:poisson}
\end{equation}
The influence of the Poisson ratio $\nu_0$ of the solid material on the elastic properties of the void material is generally not negligible. In this study $\nu_0 = 0.4$ is used, representative of many polymers, in particular of polyurethane, which is used for many foams. Only in rare cases, to allow comparison with the literature, other values are chosen.\\
The values of the stiffness and compliance tensor depend on the orientation of the corresponding basis. In order to derive the elastic properties of the material in any given direction one has to transfer the original basis  $\left\{ \mathbf{m}_1, \mathbf{m}_2, \mathbf{m}_3 \right\}$ into a new basis  $\left\{ \mathbf{e}_1, \mathbf{e}_2, \mathbf{e}_3 \right\}$ using the transformation tensor $\Omega_{ij}=\mathbf{e}_i \cdot \mathbf{m}_j$~\cite{Bower2009}. The rotation matrix $\mathbf{K}$ consists of four parts
\begin{equation}
 \mathbf{K} =  \begin{bmatrix} \mathbf{K}^{(1)} & 2 \mathbf{K}^{(2)}  \\ \mathbf{K}^{(3)} & \mathbf{K}^{(4)}  \end{bmatrix},  
\end{equation}
which can be computed from $\Omega_{ij}$ with $ i , j = 1 \ldots 3$
\begin{eqnarray}
 K_{ij}^{(1)} &=& \Omega_{ij}^2 \\
 K_{ij}^{(2)} &=& \Omega_{i\, mod(j+1,3)} \Omega_{i \, mod(j+2,3)} \\
 K_{ij}^{(3)} &=& \Omega_{mod(i+1,3) \, j} \Omega_{mod(i+2,3) \, j} \\
 K_{ij}^{(4)} &=& \Omega_{mod(i+1,3)\, mod(j+1,3)} \Omega_{mod(i+2,3)\, mod(j+2,3)} \\
                &+& \Omega_{mod(i+1,3)\, mod(j+2,3)} \Omega_{mod(i+2,3)\, mod(j+1,3)} .
\end{eqnarray}
Subsequently, one can transform the stiffness tensor according to
\begin{equation}
 \mathbf{C}^{(e)} = \mathbf{K} \, \mathbf{C}^{(m)} \,  \mathbf{K}^T. 
 \label{eq:KCK}
\end{equation}

\subsection{Finite Element Method}

To apply a certain strain $\varepsilon$ and to calculate the resulting elastic energy $U$, the commercial software ANSYS FEM was used. The RVE were meshed with tetrahedral elements with quadratic ansatz functions, controlling the mesh parameter $N$, which is the number of grid points per void spacing $L$ (see Figure~\ref{fig:fig2}). The minimum grid resolution was $N = 32$ points. In order to impose periodic boundary conditions, it is necessary to apply identical grids on opposite faces. The periodic displacement and the periodic stress is then realized by adding restricting equations to periodic point pairs, corresponding to the desired stress and strain conditions.

\subsection{Validation}
\label{sec:validation}
\begin{figure}[htbp]
	\centering
		\includegraphics[width=.5\textwidth]{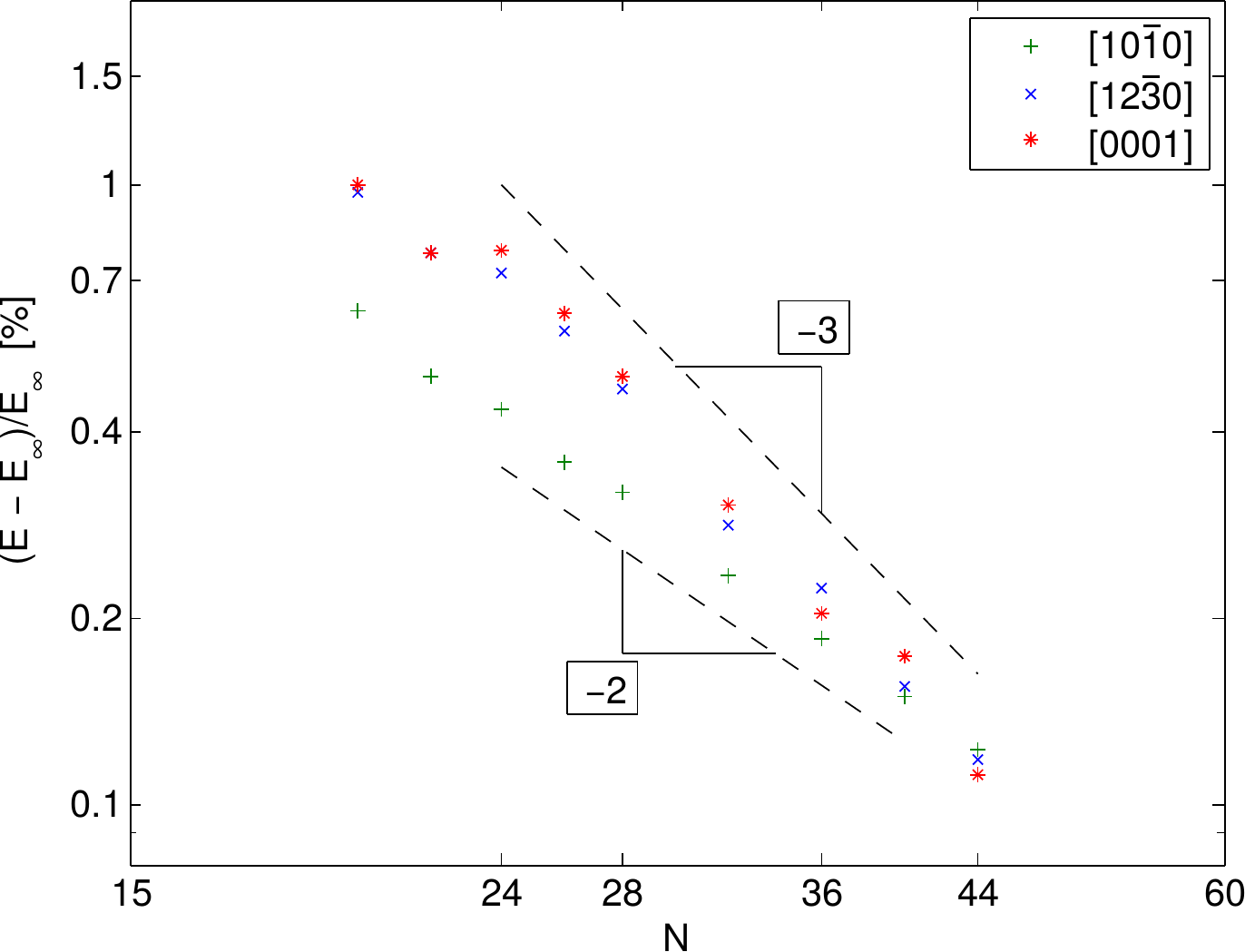} 
	\caption{ Dependence of the numerical deviation of the Young's modulus on the grid resolution $N$ for HCP at $\phi_v = 0.635$. The Young's modulus in three different directions is analysed. It is compared to the corresponding value $E_{\infty}$ for infinitely high grid resolution, derived from logarithmic fit of the computed resolutions. Broken lines show the slope for a second-order and a third-order method.}
	\label{fig:gridstudy}
\end{figure}

\begin{figure}[htb]
	\centering
		\includegraphics[width=.99\textwidth]{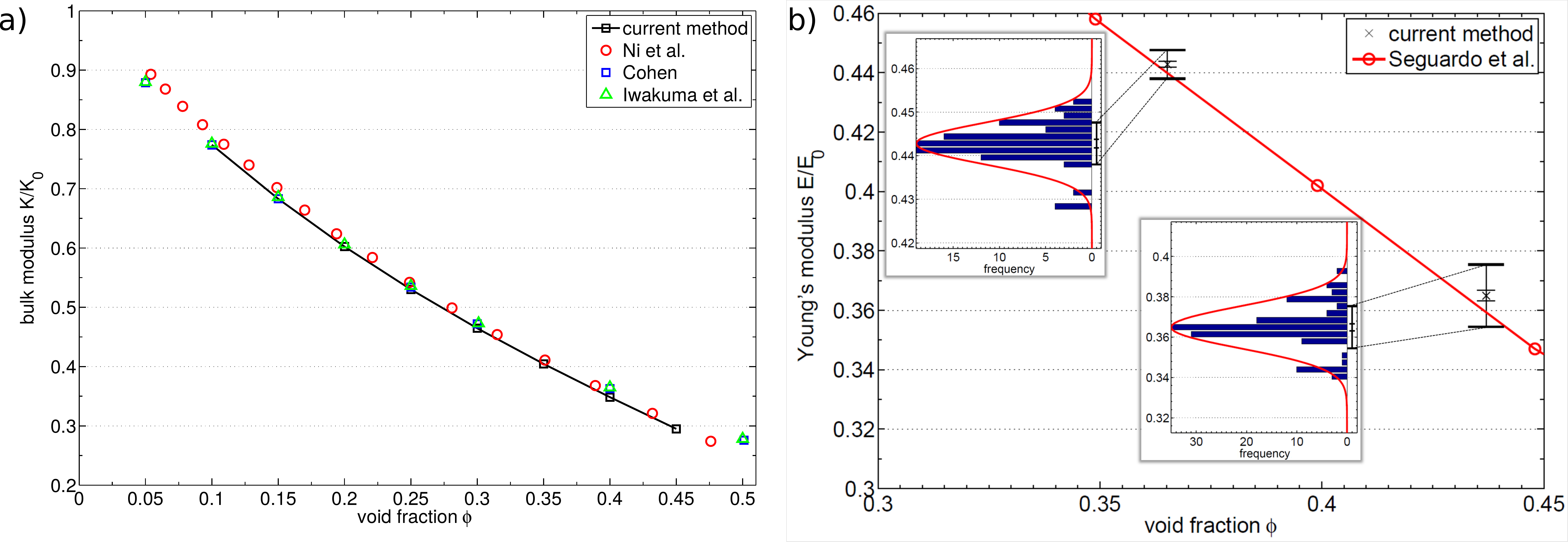}
	\caption{	Comparison of the present method with literature data. (a) Simple cubic void arrangement in \hkl [0 0 1] direction for a material with Poisson ratio $\nu_0 = 0.3$~\cite{Ni2007, Iwakuma1983, Cohen2004}. (b) Random distribution for a material with Poisson ratio $\nu_0 = 0.25$~\cite{Segurado2002}. For the present data, also the statistical variation of samples is visualised.
	Both cases show good agreement with the literature values, demonstrating the applicability of our method.}
	\label{fig:rnd_ref}
\end{figure}

\begin{figure}[htb]
	\centering
%	  	\label{fig:fig5}
		\includegraphics[width=1.0\textwidth]{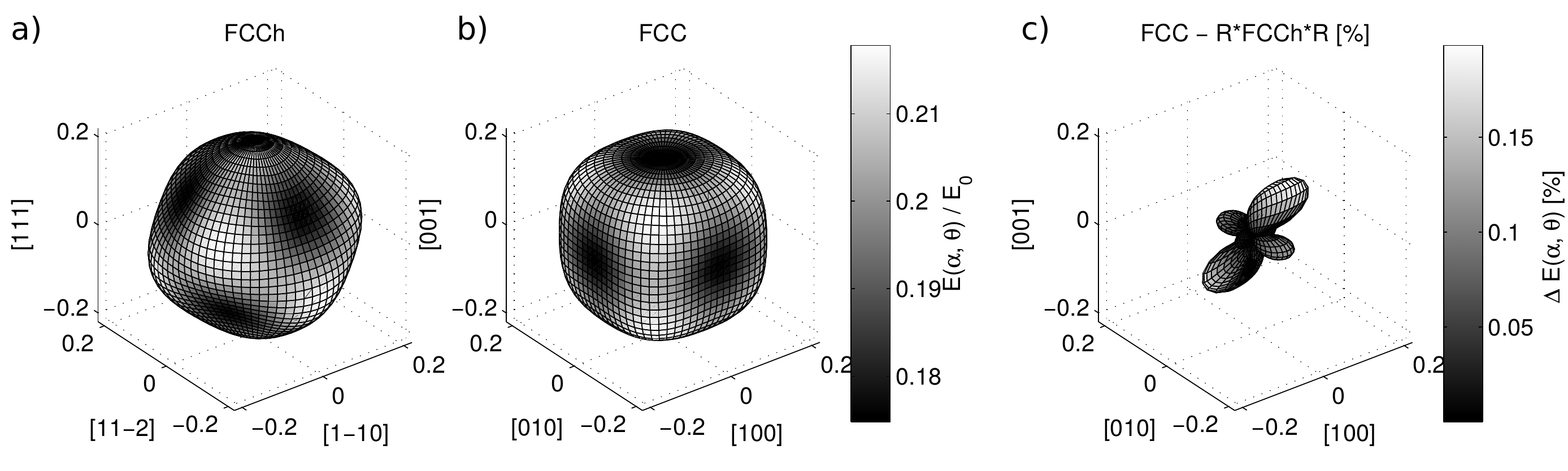} 
	\caption{Comparison of the Young's modulus, calculated by transformation of FCCh (a) or directly from FCC (b), both obtained with a resolution of $N = 44$ at a void fraction of $\phi_v = 0.64$. Colour and distance to the origin indicate the Young's modulus in the corresponding direction. (c) Percentaged difference between both Young's moduli in the corresponding direction, showing a maximum of 0.2 \%.}
	\label{fig:compare_E_rot}
\end{figure}

\begin{figure}[htbp]
	\centering
%	  	\label{fig:fig5}
		\includegraphics[width=1.0\textwidth]{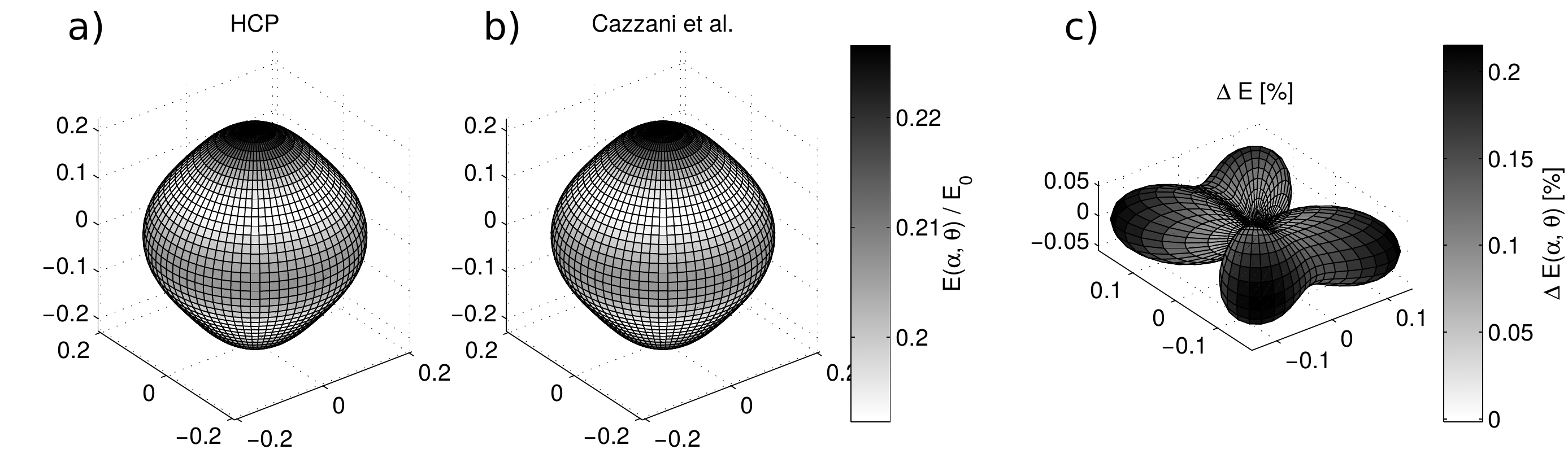} 
	\caption{Comparison of the Young's modulus for HCP, calculated from rotation of the stiffness tensor (a) and according to Cazzanis~\cite{Cazzani2014} (b), both obtained with a resolution of $N = 44$. Colour and distance to the origin indicates the Young's modulus in the corresponding direction. (c) Percentaged relative difference between the Young's moduli from a) and b) in the corresponding direction, showing a maximum of 0.2 \%.}
	\label{fig:cazzani}
\end{figure}

The results presented in this article show that differences in the mechanical behaviour occur if voids are arranged in different ways, but that some of these differences are small. In this situation it is important to assess the accuracy of the method which is applied and to demonstrate that the uncertainty of the data is below the differences addressed. Thus, three methods were applied in order to estimate the uncertainty of the obtained results.\\ 
The first method is a grid study. With increasing resolution, a numerical solution should converge toward the exact solution. However, since the resolution is usually limited due to limited computer power, one has to choose a resolution which yields results with sufficiently small deviation from the exact solution. For the grid study, HCP with thin void separation $l_l / L = 0.05$ yielding a void fraction of $0.65$ was considered. Since there is no analytical solution available for this problem, it was solved with different resolutions $20 \leq N \leq 44$. The resulting Young's modulus were fitted with a power function, yielding an approximation for the exact solution $E_{\infty}$ and the deviation $E(N)-E_{\infty}$. The results are shown in Figure~\ref{fig:gridstudy}. The method is found to be third order accurate in \hkl [1 2 -3 0] and \hkl [0 0 0 1] direction, but only second order in \hkl [1 0 -1 0] direction. The different order might arise because many of the material sheets separating two voids are oriented perpendicular to the \hkl [1 0 -1 0] direction and these material sheets are the critical regions in terms of grid resolution. A given number of grid points resolves stretching deformation better than bending deformation because of the more uniform stress and strain distribution in the stretching case. According to Figure~\ref{fig:gridstudy}, the numerical error due to resolution is well below $0.5 \%$ for $N = 32$. The results reported below were obtained with $N = 32$, except if stated otherwise. \\
The second validation method is the comparison with results to be found in literature~\cite{Segurado2002, Ni2007, Iwakuma1983, Cohen2004}. For this purpose the Poisson ratio $\nu_0$ of the solid material was chosen to be equal to the values used in the literature. Figure~\ref{fig:rnd_ref} demonstrates the good agreement between the literature data and the results of the present method. In the case of the RCP structure, simulation of 45 different, randomly generated RVEs were performed and statistically analysed. Figure~\ref{fig:rnd_ref} shows the histogram of the results, the standard deviation of the Young's modulus and the confidence interval for the mean value. The measured confidence interval equals $1\%$, which is in the same order of magnitude as the numerical uncertainty of the crystalline simulations. The statistical average value of the Young's modulus is in good agreement with~\cite{Segurado2002}.\\  
The third method of validation is an internal sensitivity test for FCCh and HCP. As pointed out in Table~\ref{tab:RVEs}, the FCC structure was calculated using two RVEs with different orientation. The resulting Young's moduli are different due to their dependence on orientation. But by applying a unitary transformation to the stiffness tensor $\mathbf{C}$, the computed result can be transformed into the same coordinate system according to Equation~(\ref{eq:KCK})
\begin{equation}
\mathbf{C'}_{\text{FCC}} = \mathbf{K} \, \mathbf{C}_{\text{FCC}} \, \mathbf{K}^{T} \label{eq:rotate}.
\end{equation}
In theory, $\mathbf{C'}_{\text{FCC}}$ and $\mathbf{C}_{\text{FCCh}}$ should be equal. But due to computational uncertainty, they exhibit a small deviation. Figure~\ref{fig:compare_E_rot} shows the difference in the Young's modulus in all directions for a gas fraction of $\phi_v = 0.64 $. This deviation yields another estimation of the uncertainty of the computation. In the present case, the maximum deviation would be below $0.8 \%$ for $N = 32$ and below $0.2 \%$ for $N = 44$. This value is in good agreement with the numerical uncertainty derived from the grid study.
\\A similar internal test can be done for HCP. Cazzani~\cite{Cazzani2014} investigated the elastic properties of materials with hexagonal close-packed structure. Taking into account the symmetries he derived an exact formula for the Young's modulus of HCP material in any direction $\mathbf{n} = (n_1, n_2, n_3)^{T}$ in terms of only four entries $D_{11}$, $D_{33}$, $D_{66}$, and $D_{31}$ of the compliance tensor
\begin{equation}
 E(\mathbf{n}) = \left\{  D_{33} - \left[ \left( D_{33} - D_{11} \right) n_1^2 + \left( 2 D_{33} - 2 D_{31} - D_{66} \right) \left( n_2^2 + n_3^2 \right) \right] n_1^2 \right\} ^{-1} .
 \label{eq:cazzani}
\end{equation}
As explained above, one may calculate the Young's modulus in any direction by rotating the stiffness tensor according to Equation~(\ref{eq:KCK}). This yields two independent ways to calculate the angular distribution of the Young's modulus. The resulting values of both methods can be compared, as demonstrated in Figure~\ref{fig:cazzani}. 
The maximum deviation is below $0.2 \%$ for $N = 44$, which is again in line with the findings above.\\
Taking into account the different tests performed on the uncertainty of the method proves that overall the uncertainty of the employed numerical method for the computation of the Young's modulus is less than $1 \%$ for $N = 32$ and less than $0.2 \%$ for $N = 44$.

\afterpage{\clearpage}
\section{Results and Discussion}
\label{sec:results_and_discussion}

\subsection{Young's Modulus}

The Young's modulus for the different configurations given in Table~\ref{tab:RVEs} was computed for a range of void fractions $0 \leq \phi_v \leq 0.95$. The void fraction of the RVE was varied by changing the sphere diameter $D$ while fixing the bubble centre positions. The actual void fraction was calculated from the volume of all finite elements after meshing. The method also allows for sphere diameters larger than the distances between the bubbles, yielding overlapping voids and thus, giving rise to open-cell structures. The mean values of the Young's modulus $E_{mean}$ (defined in Equation~\ref{eq:Emean} below) are plotted in Figure~\ref{fig:E_mean}. 
\begin{figure}[htb]
	\centering
	  \includegraphics[width=.7\textwidth]{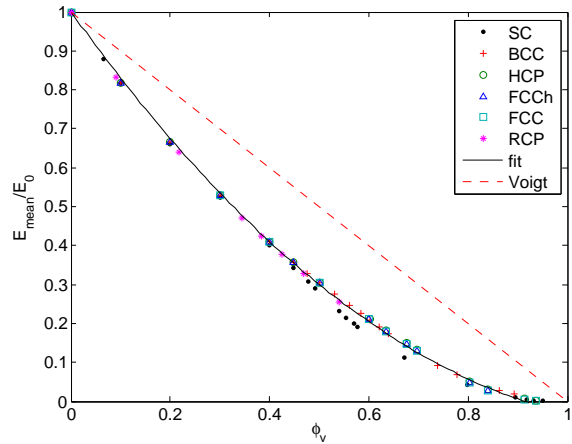} 
	\caption{Mean Young's modulus (Equation~(\ref{eq:Emean})) over a range of void fractions $\phi_v$ for different arrangements of spherical voids. The solid line represents the fitting curve from Equation~(\ref{eq:E_fit}) while the broken line represents Voigt's upper bound. }
	\label{fig:E_mean}
\end{figure}
Generally, the mean Young's moduli of different structures appear to be relatively close to each other and well below Voigt's bound (Equation~(\ref{eq:voigt})). A rough estimation of the mean Young's modulus of void material for a given void fraction can be extracted by least-square fitting of the data in Figure~\ref{fig:E_mean}, yielding
\begin{equation}
 E_{mean}(\phi_v) \approx 0.74 \phi_v^2 - 1.77 \phi_v +1.
 \label{eq:E_fit}
\end{equation}
This fit implies a quadratic dependence of the mean Young's modulus on the void fraction. It goes to zero for void fractions of 0.915, which is well below the range of rigidity-loss above $\phi_v = 0.99$. This shows that for open-cell structures at high void fractions the quadratic scaling does not represent the actual dependence very well. In an associated study~\cite{Heitkam2016} we have shown that the Young's modulus in the limit of rigidity-loss scales with the solid fraction to the power of 3.5.
\\The structural differences manifest themselves mostly in the anisotropy which will be addressed now. To that end, the corresponding stiffness tensor $C$ for each structure and void fraction was rotated according to Equation~(\ref{eq:KCK}) from the original basis $\left\{ \mathbf{m}_1, \mathbf{m}_2, \mathbf{m}_3 \right\}$ into the new orthonormal basis $\left\{ \mathbf{e}_1, \mathbf{e}_2, \mathbf{e}_3 \right\}$ %(Figure~\ref{fig:sketch_e123_E})
\begin{equation}
 \mathbf{e}_1 =  \begin{pmatrix} \sin(\theta) \cos(\alpha) \\ \sin(\theta) \sin(\alpha) \\ \cos(\theta) \end{pmatrix},   \mathbf{e}_2 = \begin{pmatrix} -\sin(\alpha) \\ \cos(\alpha) \\ 0 \end{pmatrix},  \mathbf{e}_3 = \begin{pmatrix} - \cos(\theta) \cos(\alpha) \\ - \cos(\theta) \sin(\alpha) \\ \sin(\theta) \end{pmatrix},   
\end{equation}
varying $\alpha \in [0, 2\pi [$ and $\theta \in [0, \pi ]$. Figure~\ref{fig:E_winkel_kombi} shows the dependence of $E_{11}$ on the angles $\alpha$ and $\theta$ for selected void fractions.  
\begin{figure}[htbp]
	\centering
	  \includegraphics[width=.9\textwidth]{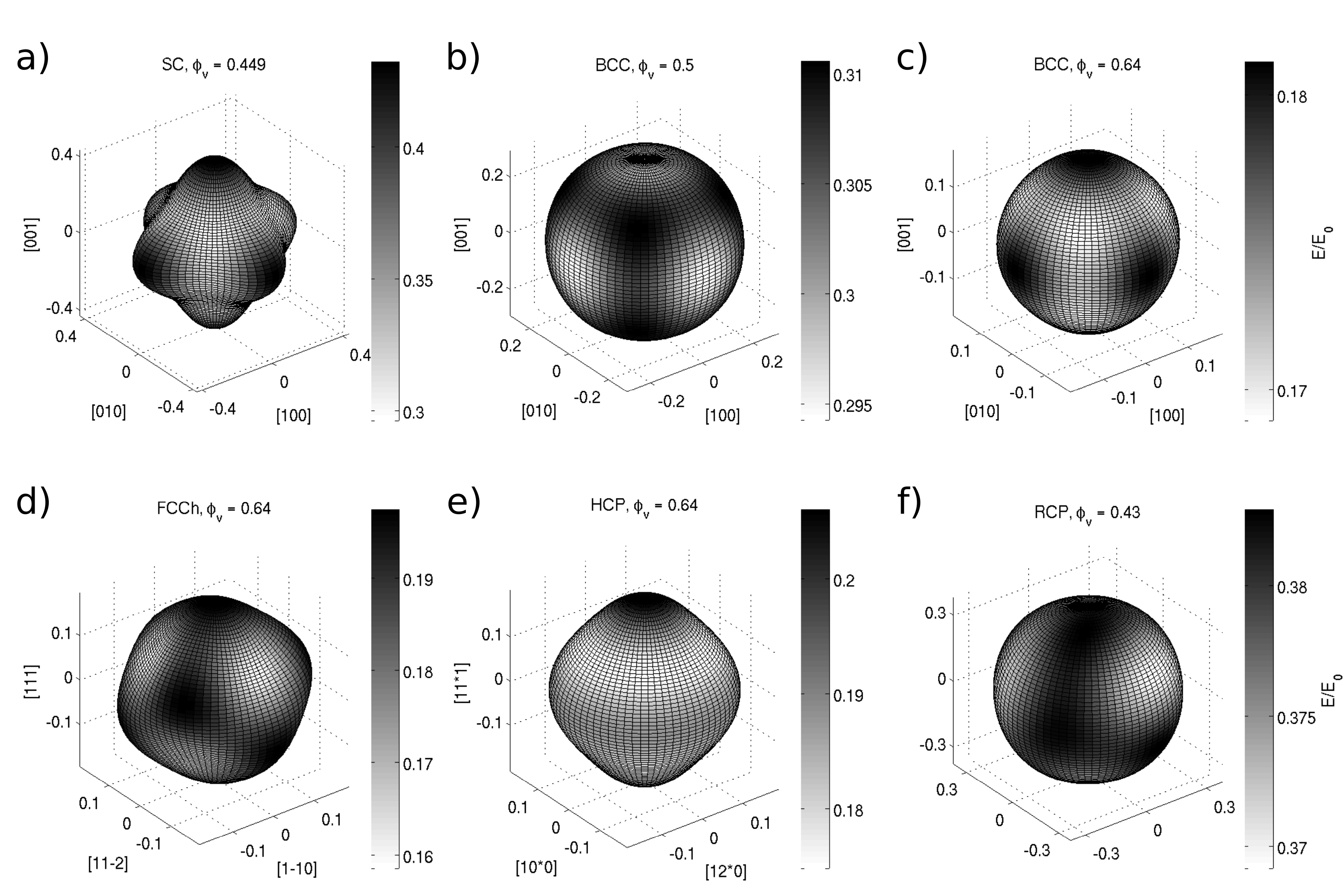} 
	\caption{Dependence of the Young's modulus on the direction for different void arrangement at certain void fractions $\phi_v$. Colour represents the relative Young's modulus. Distance from the origin also represents the Young's modulus in the corresponding direction $r(\alpha, \theta) = E_{11}(\alpha, \theta, \phi_v)$}
	\label{fig:E_winkel_kombi}
\end{figure}
Cubic void arrangements, such as SC, BCC and FCCh show a cubic symmetry in the angular dependence of the Young's modulus. HCP schows isotropic dependence around the \hkl [0 0 0 1] direction. For RCP only small deviation from full isotropy is found, corresponding to the finite number of voids in an RVE of RCP.
The maximum value, $E_{max}$, the minimum value, $E_{min}$, and the mean value, $E_{mean}$ of $E_{11}$ are extracted by 
\begin{eqnarray}
           E_{max} (\phi_v) &=&  \mathrm{max}_{\forall (\alpha, \theta) \in  [0,2 \pi[ \times [0, \pi]} E_{11}(\alpha, \theta, \phi_v) \nonumber  \\
           E_{min} (\phi_v) &=&  \mathrm{min}_{\forall (\alpha, \theta) \in  [0,2 \pi[ \times [0, \pi]} E_{11}(\alpha, \theta, \phi_v) \nonumber \\  
           E_{mean}(\phi_v) &=& \frac{1}{4 \pi} \int_0^{\pi} \int_0^{2 \pi}  E_{11}(\alpha, \theta, \phi_v)  \sin(\theta) d\alpha d\theta . 
           \label{eq:Emean}
\end{eqnarray}
Figure~\ref{fig:E_max_min} shows the maximum and minimum Young's modulus for different void arrangements at different void fractions. In order to distinguish the structures more clearly, the figure displays the difference between the Young's modulus and the Voigt's bound 
\begin{equation}
\frac{E(\phi_v) - E_{Voigt} (\phi_v)}{E_0}  = \frac{E(\phi_v)}{E_0} - \phi_s .
\label{eq:rel_def}
\end{equation}
\begin{figure}[htbp]
	\centering
	  \includegraphics[width=1.0\textwidth]{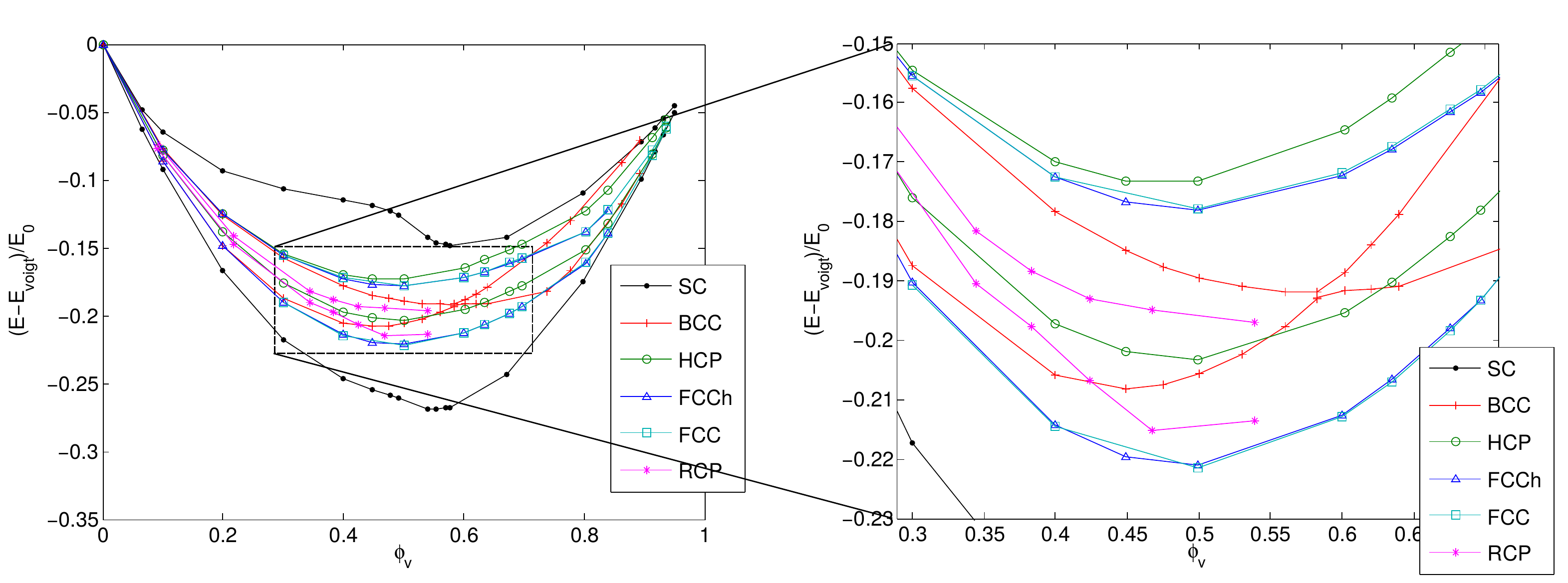} 
	\caption{Difference of the Young's modulus to the Voigt's bound for a range of void fractions $\phi_v$. For each void structure the maximum and minimum Young's modulus is shown, derived from transformation of the stiffness tensor according to Equation~(\ref{eq:Emean}). The figure compares all configurations investigated.}
	\label{fig:E_max_min}
\end{figure}
Simple cubic shows very prominent maximum and minimum values of the Young's modulus. All the other values are close to each other. Apart from SC, the transition from closed-cell to open-cell void material seems to change the Young's modulus only slightly. FCCh and FCC show nearly perfect agreement, as already observed in the validation section.

\afterpage{\clearpage}
\subsection{Poisson ratio}

\begin{figure}[htbp]
	\centering
	  \includegraphics[width=.95\textwidth]{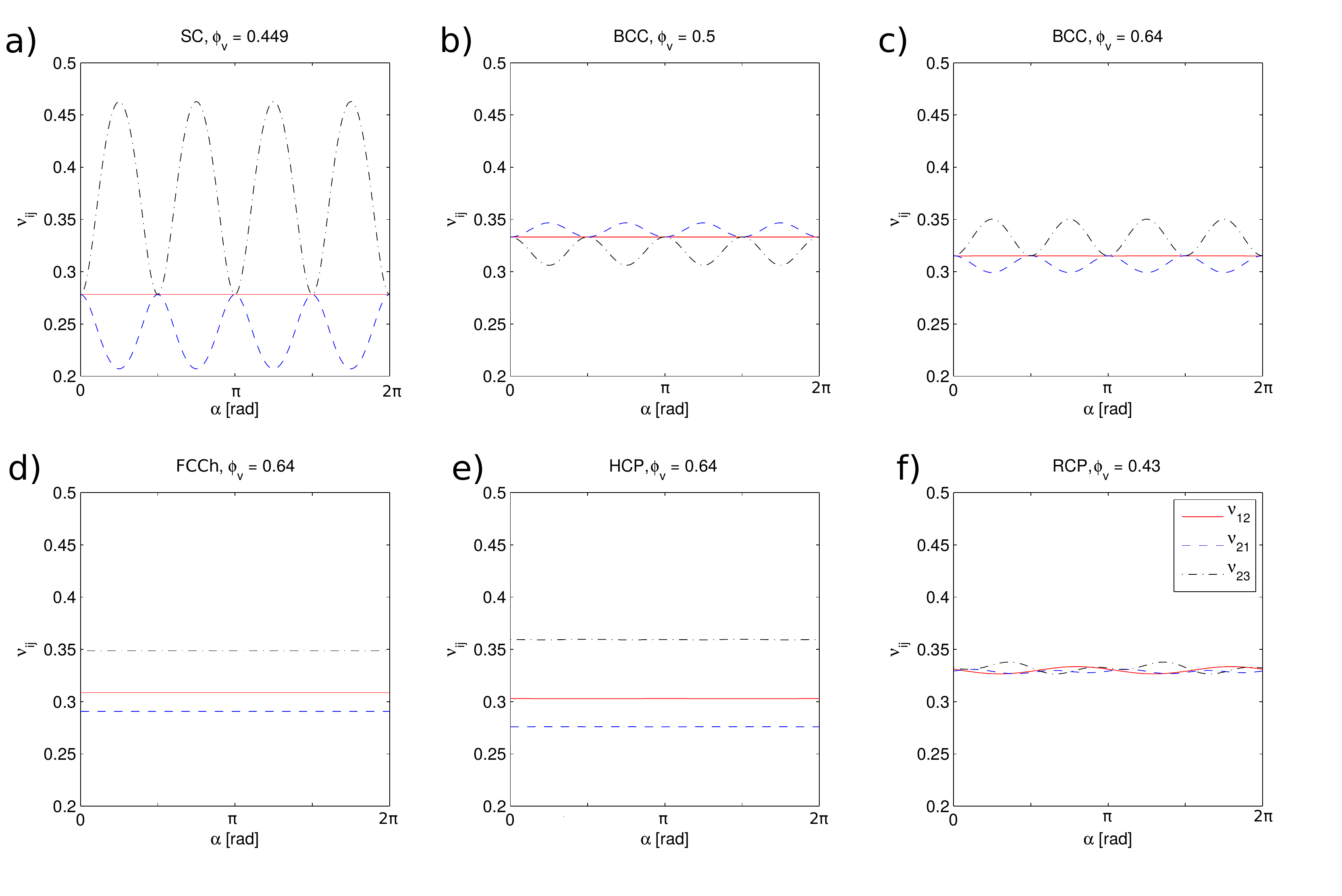} 
	\caption{Dependence of the Poisson ratios $\nu_{12}$, $\nu_{23}$, and $\nu_{21}$ on the direction $\alpha$ of $\mathbf{f}_2$ for different void arrangements at certain void fractions.}
	\label{fig:nu_winkel_kombi}
\end{figure}

\begin{figure}[htbp]
	\centering
	  \includegraphics[width=.99\textwidth]{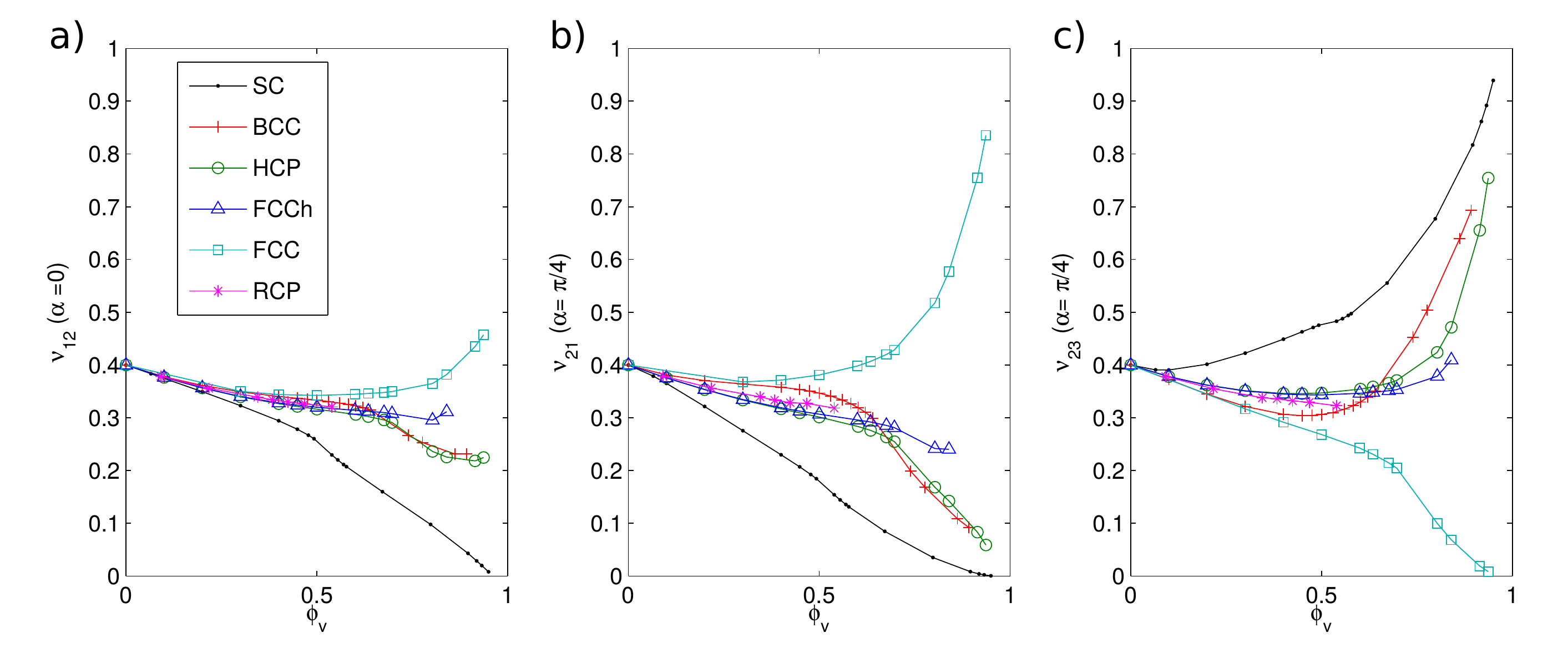} 
	  \caption{Poisson ratios for a range of void fractions $\phi_v$ for different arrangements of spherical voids. Each plot shows one of the selected Poisson ratios $\nu_{12}(\alpha = 0)$, $\nu_{23}(\alpha = \pi / 4)$, and $\nu_{21}(\alpha = \pi / 4)$.  They will be discussed in separate subsections below.}
	\label{fig:nu_max_min}
\end{figure}

From the stiffness tensor the Poisson ratio for any pair of orthogonal directions $\mathbf{f}_i$ and $\mathbf{f}_j$ can be derived by rotation of the basis of the tensor according to Equation~\eqref{eq:rotate} and subsequent application of Equation~\eqref{eq:poisson}. In this study, only special combinations of directions are investigated, corresponding to the symmetries of the void arrangements. To that end, a new orthonormal basis $\left\{ \mathbf{f}_1, \mathbf{f}_2, \mathbf{f}_3 \right\}$ was defined with respect to the original basis $\left\{ \mathbf{m}_1, \mathbf{m}_2, \mathbf{m}_3 \right\}$
\begin{equation}
 \mathbf{f}_1 =  \begin{pmatrix} 0 \\ 0 \\ 1 \end{pmatrix},   \mathbf{f}_2 = \begin{pmatrix} \cos(\alpha) \\ \sin(\alpha) \\ 0 \end{pmatrix},  \mathbf{f}_3 = \begin{pmatrix} -\sin(\alpha) \\ \cos(\alpha) \\ 0 \end{pmatrix}.   
\end{equation}
The direction $\mathbf{f}_1$ is fixed, parallel to the RVE axis $\mathbf{m}_3$, defined in Figure~\ref{fig:RVE}. The angle $\alpha \in [0, 2\pi [$ was varied, so that $\mathbf{f}_2$ and $\mathbf{f}_3$ cover the complete plane, orthogonal to $\mathbf{f}_1$. The corresponding Poisson ratios $\nu_{12}$, $\nu_{23}$ and $\nu_{21}$ are given in Figure~\ref{fig:nu_winkel_kombi}.\\
For cubic void arrangements, such as SC and BCC, four-fold symmetry of $\nu_{21}$ and $\nu_{23}$ around the $\mathbf{f}_1$ direction is visible, referring to the four-fold symmetry of a cube. For $\alpha = k \pi/2, k \in \mathbb{N}$ the basis $\left\{ \mathbf{f}_1, \mathbf{f}_2, \mathbf{f}_3 \right\}$ coincides with the cubic axes, so that $\nu_{12}$, $\nu_{21}$, and $\nu_{23}$ are equal. The Poisson ratio $\nu_{12}$ is independent of the direction of $\mathbf{f}_2$, because four-fold symmetry in the $\{\mathbf{f}_2,\mathbf{f}_3\}$ layer is sufficient to assume isotropy of the resulting strain in this layer. The hexagonal structures FCCh and HCP show three-fold symmetry in the $\{\mathbf{f}_2,\mathbf{f}_3\}$ layer causing $\nu_{12}$ $\nu_{21}$, and $\nu_{23}$ to be independent of the direction of $\mathbf{f}_2$. For RCP only small deviations from isotropy are visible, resulting from the finite number of voids in the RVE.
\\From these angular dependencies, three characteristic combinations of directions are derived. These are $\nu_{12}(\alpha = 0)$, $\nu_{23}(\alpha = \pi / 4)$, and $\nu_{21}(\alpha = \pi / 4)$. Figure~\ref{fig:nu_max_min} shows the dependence of these selected Poisson ratios on the void fraction and on the void arrangement. 
For vanishing void fraction all Poisson ratios converge to $\nu = 0.4$ which is the Poisson ratio of the matrix material. It is interesting to note, that for high void fractions Poisson ratios higher than $0.5$ appear, which is not possible for isotropic material but can be the case for anisotropic materials. 
For isotropic material, a Poisson ratio higher than $0.5$ means, that the volume increases under uni-axial compression, which implies a negative bulk modulus. For anisotropic void material high Poisson ratios are possible. The issue was investigated by Ting et al.~\cite{Ting2005} in the framework of general anisotropic material. In zinc~\cite{Lubarda1999} and cubic metals~\cite{Baughman1998} researchers found negative Poisson ratios. However, in the present study all computed Poisson ratios are positive. But, for high void fractions close to rigidity-loss Poisson ratios close to zero appear in some cases.\\ 
In the following, the different structures are discussed separately, concerning Young's modulus and Poisson ratio.

\afterpage{\clearpage}
\subsection{Simple cubic}

\begin{figure}[htbp]
	\centering
	  \includegraphics[width=.75\textwidth]{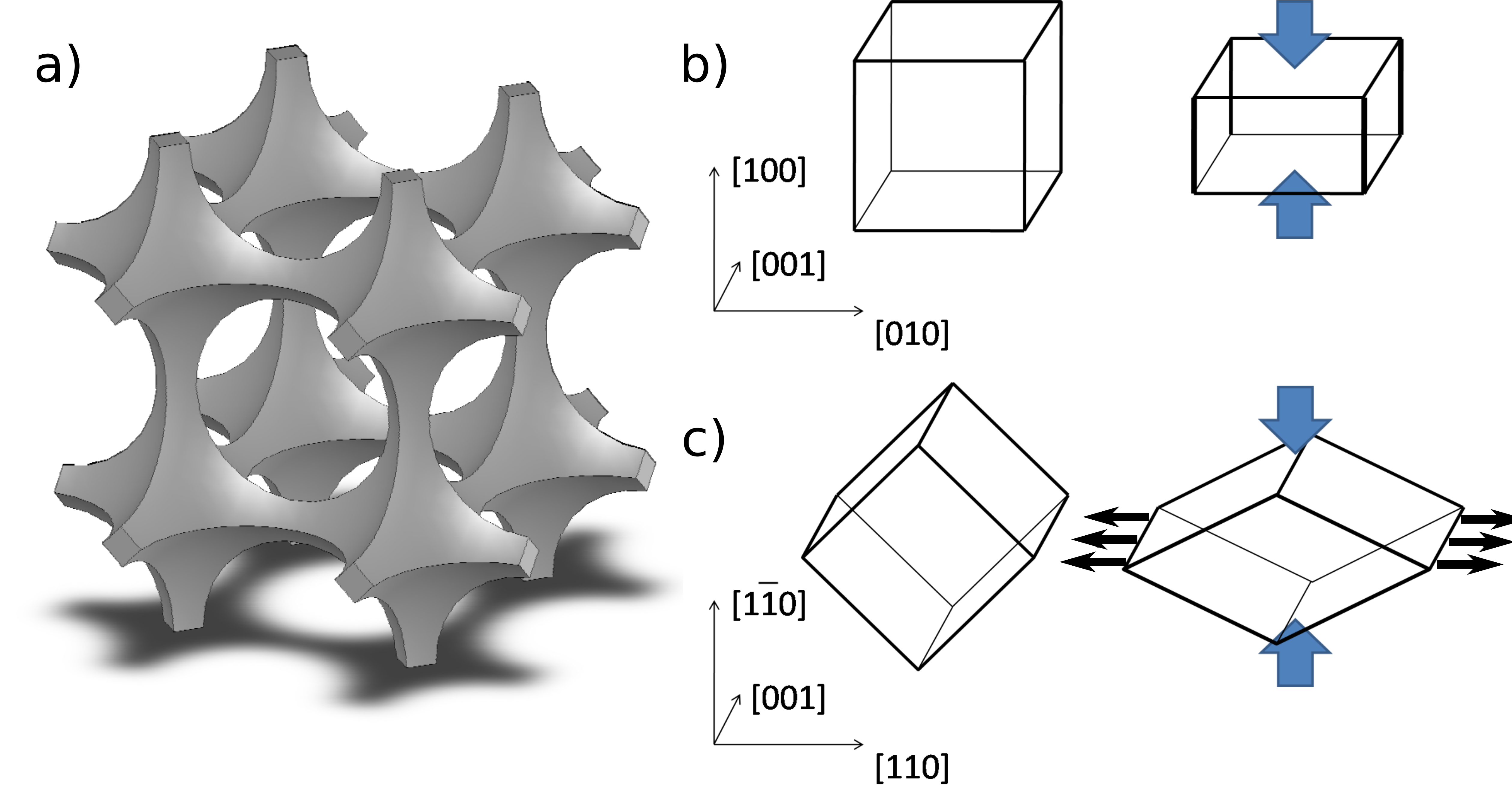} 
	\caption{(a) Structure of the interstitial material for simple cubic arrangement of overlapping spherical voids at $\phi_v = 0.75$. The structure corresponds to beams forming a cubic wire-frame. (b) Mechanism of deformation when straining a cubic wire-frame. Strain along the \hkl [1 0 0] direction causes perpendicular relaxation of the individual wires but no relaxation of the structure. (c) Strain along the \hkl [1 -1 0] direction causes perpendicular relaxation of the structure in the \hkl [1 1 0] direction.}
	\label{fig:sc_kombi1}
\end{figure}

%The Young's modulus for the simple cubic void arrangement is shown again in Figure~\ref{fig:E_max_min}. 
%
Compared to other void arrangements the simple cubic one shows very disparate maximum and minimum values in the Young's modulus. In Figure~\ref{fig:E_winkel_kombi}a the Young's modulus in different directions is plotted for a void fraction of $\phi_v = 0.449$, corresponding to thin material sheets between the voids. It shows a clear maximum in the \hkl [1 0 0], \hkl [0 1 0] and \hkl [0 0 1] directions.\\  
The reason for this high Young's modulus can be inferred from consideration of the structure of the interstitial material shown in Figure~\ref{fig:sc_kombi1}a. The material forms columns in the \hkl [1 0 0] direction, supporting a load very effectively, as illustrated in Figure~\ref{fig:sc_kombi1}b. But applying the load at an angle to the column direction, e.g. in \hkl [1 -1 0] direction (Figure~\ref{fig:sc_kombi1}c), this causes shearing of the structure. Therefore, the Young's modulus shows a minimum in \hkl [1 1 1] direction. \\
The extremal values of the Poisson ratio and their dependence on the void fraction is shown in Figure~\ref{fig:nu_max_min}. In Figure~\ref{fig:nu_winkel_kombi} the dependence of the Poisson ratio on the direction is presented for a void fraction of $\phi_v = 0.449$. The Poisson ratio $\nu_{12}$ for compression in \hkl [1 0 0] direction is small and decreases with increasing void fraction. This is also due to the column-like structure in this direction (Figure~\ref{fig:sc_kombi1}a). Compression of the structure causes relaxation of the individual columns in transversal direction, but not relaxation of the whole structure (Figure~\ref{fig:sc_kombi1}b). Buckling of the column is not relevant to the linear-elastic simulation.\\
In contrast to the \hkl [1 0 0] direction, stress along the \hkl [1 -1 0] direction is distributed on columns in \hkl [1 0 0] and \hkl [0 1 0] direction. This causes very strong relaxation in the \hkl [1 1 0] direction because the structure is folded (Figure~\ref{fig:sc_kombi1}c). Therefore, the maximum value of $\nu_{23}$ in Figure~\ref{fig:nu_max_min} increases significantly with decreasing void fraction. For values of $\alpha$ that give a maximum of $\nu_{23}$ the corresponding $\nu_{21}$ shows a minimum (Figure~\ref{fig:nu_winkel_kombi}).\\
Below a void fraction of $\phi_{v,touch} = 0.524$ the voids in SC arrangement are separated, forming a closed-cell structure. Above this void fraction, the voids cut through each other (Figure~\ref{fig:sc_kombi1}a). Crossing this critical void fraction, the Young's modulus shows a smooth reduction by about 10 \% (Figure~\ref{fig:E_max_min}), because the material sheets between voids are removed. Also the Poisson ratios (Figure~\ref{fig:nu_max_min}) show a significant reduction crossing the void fraction $\phi_{v,touch}$. In contrast to other void arrangements, many of these material sheets are oriented parallel to the cubic axes, adding substantially to the Young's modulus along the cubic axes because these sheets act the same way as the columns, supporting an external load by elongation. Also, the material sheets connect the columns and stiffen the structure when the load is not applied along the column direction. \\

\afterpage{\clearpage}
\subsection{Body centred cubic}
\begin{figure}[htbp]
	\centering
	  \includegraphics[width=.7\textwidth]{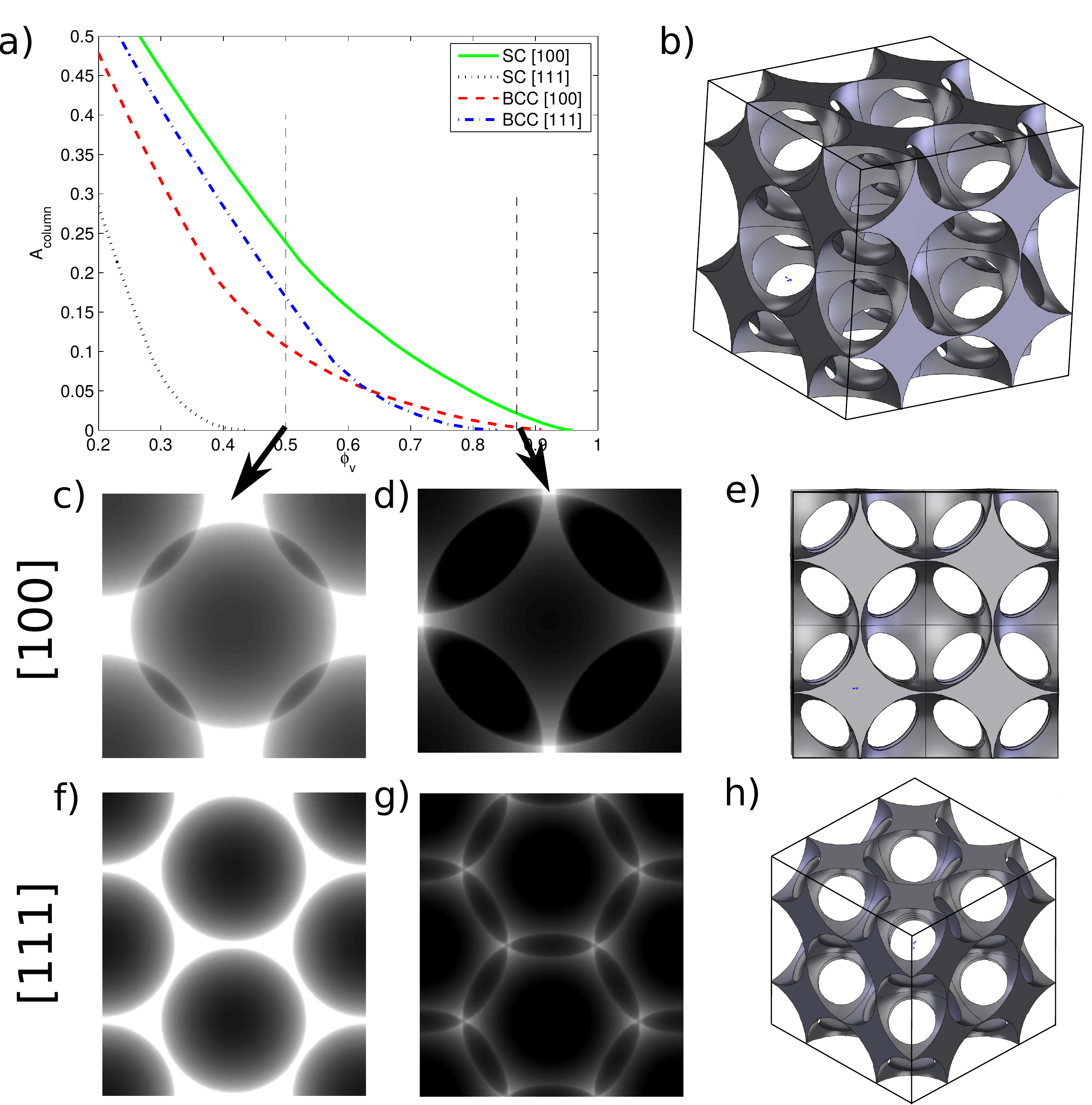} 
	\caption{Dependence of the column thickness on void fraction, structure and orientation. (a) Fraction of the cross section forming an unbroken column for BCC and SC. (b,e,h) Visualisation of the BCC void material at $\phi_v = 0.94$ under different angles. (c,d,f,g) Distribution of material in the case of BCC, projected in \hkl [1 0 0] (c,d,e) and \hkl [1 1 1] direction (f,g,h) at $\phi_v = 0.54$ (c,f) and $\phi_v = 0.94$ (d,g), respectively. White colour corresponds to values of 1 and marks unbroken columns.}
	\label{fig:compare_columns}
\end{figure}

The Young's modulus for the body centred cubic void arrangement (Figure~\ref{fig:E_max_min}) shows a very interesting behaviour, in that at a void fraction of $\phi_v \approx 0.59$ the directions for maximum and minimum Young's modulus switch. At this point the Young's modulus is the same for all directions. Figure~\ref{fig:nu_winkel_kombi} shows, that also the Poisson ratios all equal 0.32 at this void fraction. That Young's modulus and Poisson ratio are independent of the angle means, that BCC at a void fraction of 0.59 is elastically isotropic. Figure~\ref{fig:E_winkel_kombi} shows the angular dependence of the Young's modulus above and below this switching point, respectively. For lower void fraction, the maximum is oriented in \hkl [1 1 1] direction, for higher void fraction in \hkl [1 0 0] direction. BCC with high void fraction shows an angular distribution of the Young's modulus similar to SC with maximum Young's modulus along the columns in the cubic axes.\\
Figure~\ref{fig:compare_columns} shows the structure of the interstitial material. Again, the elastic behaviour can be explained with the column structure. Figure~\ref{fig:compare_columns} illustrates the existence of columns. Figure~\ref{fig:compare_columns}a gives the fraction of the cross section that forms an unbroken column for different structures, orientations and void fractions. To derive the cross section of unbroken columns, the three-dimensional distribution of solid fraction of the void material is projected in a direction parallel to the columns, yielding the two-dimensional distribution of material. This two-dimensional distribution is shown in Figure~\ref{fig:compare_columns}c,d,f,g for BCC at different void fractions and different directions of projection. For interpretation, they can be thought of as idealised X-ray photographies of the structure. The share of the area that equals 1 in the two-dimensional distribution represents the cross sectional area of an unbroken column in the direction of projection. In case of a body centred cubic arrangement this cross-sectional area behaves similar to the Young's modulus. For void fractions below $0.6$ the larger area is oriented in \hkl [1 1 1] direction. Above $\phi_v = 0.6$ the \hkl [1 0 0] orientation contains stronger columns. The value of $\phi_v = 0.6$ coincides with the void fraction at which the elastic behaviour of BCC switches. At $\phi_v=0.84$ the straight columns in \hkl [1 1 1] direction completely disappear. The columns in \hkl [1 0 0] direction remain up to void fractions of $\phi_v = 0.94$.\\ 
The extremal values of the Poisson ratio and their dependence on the angle $\alpha$ is shown in Figure~\ref{fig:nu_max_min} and~\ref{fig:nu_winkel_kombi}, respectively. The latter is performed for a void fraction of $\phi_v = 0.5$ and $\phi_v = 0.64$. The Poisson ratio also shows a switching of behaviour at $\phi_v \approx 0.59$. Again, BCC with high void fraction behaves qualitatively similar to SC, with high values of $\nu_{23}$ and low values of $\nu_{21}$. Again, this can be explained with the orientation of columns in the void material. For high void fraction, columns are oriented along the cubic axes (similar to SC). Stress along the cubic axes is supported by these columns, yielding low relaxation in the perpendicular direction. Stress diagonal to the column direction, e.g. along the \hkl [1 -1 0] direction, is distributed to the \hkl [1 0 0] and \hkl [0 1 0] column, which causes folding of the structure in \hkl [-1 1 0] direction, as sketched in Figure~\ref{fig:sc_kombi1}b. This results in high values of $\nu_{23}$ but even lower values of $\nu_{21}$. BCC at low void fractions behaves completely different. The strongest columns are oriented in \hkl [1 1 1], \hkl [-1 1 1], \hkl [1 -1 1], and \hkl [1 1 -1] direction. Stress along the \hkl [1 -1 0] direction, hence, is distributed on the columns in \hkl [1 -1 1] and \hkl [-1 1 1]. This causes a folding of the structure in \hkl [0 0 1] direction, yielding high values of $\nu_{21}$ but low values of $\nu_{23}$.

\afterpage{\clearpage}
\subsection{Face centred cubic and hexagonal close packed}
\begin{figure}[htbp]
 	\centering
 		\includegraphics[width=.7\textwidth]{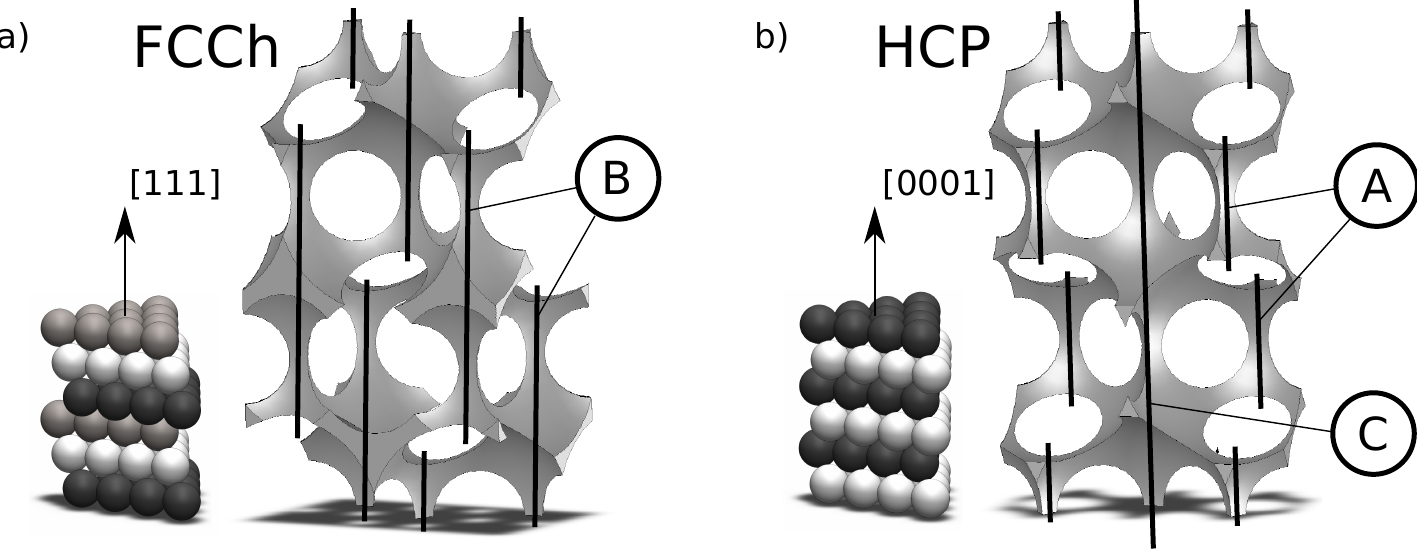} 
 	\caption{Structure of (a) FCCh and (b) HCP void arrangement in hexagonally ordered layers. In the visualization of the column structure of the interstitial material obtained with spherical voids for $\phi_v = 0.85$ FCCh shows broken columns B while HCP shows unbroken columns C and broken side-columns A. }
 	\label{fig:hcp_fcc_column}
 \end{figure}

In order to compare FCC and HCP in a direct way, the hexagonal orientation of FCC, here called FCCh, is considered. As shown in Figure~\ref{fig:RVE}, the RVE of FCCh is bounded by planes in \hkl (1 0 0), \hkl (0 1 -1) and \hkl (1 1 1) orientation. The two void arrangements, FCCh and HCP, have a very similar structure. The voids are placed in layers of hexagonal ordering. These layers are stacked in different sequences, shown in Figure~\ref{fig:hcp_fcc_column}.\\
For both structures, the dependence of the Young's modulus on the void fraction an on the direction is shown in Figure~\ref{fig:E_max_min} and Figure~\ref{fig:E_winkel_kombi}. 
The HCP arrangement shows a clear maximum in the \hkl [0 0 0 1] direction. This effect is similar as observed with the simple cubic arrangement. In this direction, the structure of the interstitial material contains a straight column, marked C in Figure~\ref{fig:hcp_fcc_column}b, which supports the load very efficiently. HCP also contains broken side-columns A that contribute only slightly to the Young's modulus in \hkl [0 0 0 1] direction. 
\\FCCh shows four maxima in the \hkl [1 1 1], \hkl [1 1 -1], \hkl [1 -1 1] and \hkl [-1 1 1] direction, which is consistent with its cubic symmetry. In these directions, the FCCh structure contains columns that are broken in every third layer, marked B in Figure~\ref{fig:hcp_fcc_column}a. As a result, the maximal Young's moduli for FCCh are a bit smaller than the maximum for HCP. To support load along these broken columns, the load has to be transferred between the columns. This is done by shearing the material sheets between the columns. For higher void fraction, the material sheets become very thin so that load transfer between the broken columns causes stronger shearing deformation. Supporting the load along the unbroken column C of HCP does not rely on shearing of material sheets. Therefore, the maximum Young's modulus of FCCh decreases faster than the maximum of HCP with increasing void fraction, as shown in Figure~\ref{fig:E_max_min}.\\
 \begin{figure}[htbp]
 	\centering
 		\includegraphics[width=.7\textwidth]{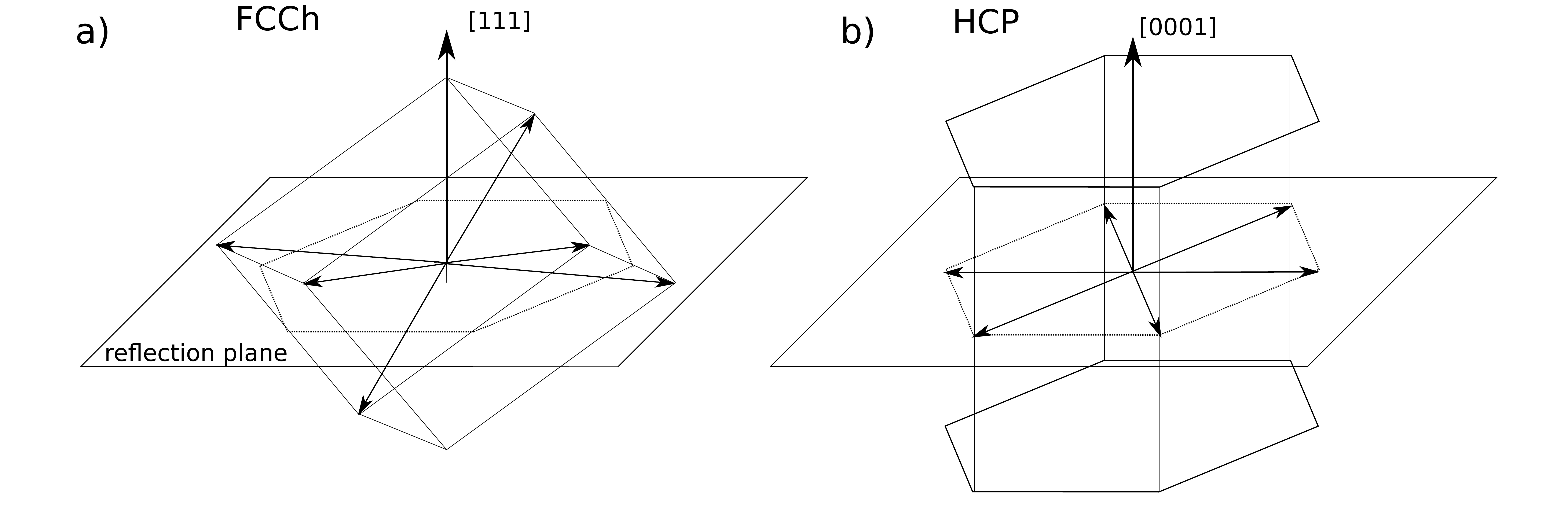} 
 	\caption{Symmetry in FCCh and HCP sphere packing. (a) FCCh showing only three-fold symmetry around the \hkl [1 1 1] axis, but including reflection on the reflection plane, six-fold symmetry holds. (b) HCP shows six-fold symmetry around the \hkl [0 0 0 1] axis. }
 	\label{fig:symm}
 \end{figure}
The Young's modulus of HCP appears to be symmetric around the \hkl [0 0 0 1] axis while for FCCh it shows cubic symmetry. However, in the \hkl (0 0 0 1) and \hkl (1 1 1) plane, respectively, the Young's moduli of HCP and FCC are isotropic (Figure~\ref{fig:E_winkel_kombi}). The isotropy of FCCh and HCP in the basal plane can be related to crystal symmetry as illustrated in Figure~\ref{fig:symm}. Macroscopic properties such as elastic moduli must conform to the external symmetry of the crystal. In the case of HCP, this includes a six-fold rotation about the \hkl [0 0 0 1] axis, perpendicular to the hexagonal layers. A well known theorem states that this is sufficient to ensure that the stiffness tensor (which is fourth rank) is transversely isotropic~\cite{Gibson1997}, precisely the property exhibited in the calculations (Figure~\ref{fig:E_winkel_kombi}). Also Equation~(\ref{eq:cazzani}) derived by Cazzani~\cite{Cazzani2014} reflects this isotropy, yielding a dependence of the Young's modulus on $(n_2^2 + n_3^2)$.\\
In the case of FCCh, transverse isotropy is not observed, but at least Young's modulus is isotropic for imposed strain in the \hkl (1 1 1) plane (and also in the \hkl (-1 1 1), \hkl (1 -1 1), and \hkl (1 1 -1) plane)  as observed in Figure~\ref{fig:E_winkel_kombi}. For FCC there is indeed no six-fold rotation included in its point group. Instead, the axes \hkl [-1 1 1], \hkl [1 -1 1], and \hkl [1 1 -1] correspond to a three-fold symmetry around the \hkl [1 1 1] axis. Three-fold symmetry is not sufficient to ensure isotropy. Including a reflection at the \hkl [1 1 1] plane three-fold symmetry is transformed into six-fold symmetry. The reflection-based six-fold symmetry implies isotropy only if the direction of strain is not changed by the reflection. Apart from the axis itself, the only directions for strain which are not altered by this reflection are those which lie in the reflection plane. Hence, a limited isotropy holds for directions of strain within such a reflection plane only, as observed.\\
Figure~\ref{fig:hcp_fccH_iso} shows the dependence of the Young's modulus on the angle $\alpha$ for $\theta = 0$. Perfect isotropy is not reproduced by the numerical method. The fluctuations are below 0.2 \% which is in line with the numerical uncertainty of the method. Additionally, FCCh and HCP exhibit different values for the Young's modulus in the basal plane. This is surprising, since both structures consist of layers of hexagonally arranged voids. The difference in the Young's modulus between FCCh and HCP parallel to these layers equals about $1 \%$ for increased numerical resolution $N=50$. Since the numerical uncertainty for the increased resolution was found to be smaller than $0.2 \%$, the difference in the Young's modulus is definitely physical. The interstitial material of both void structures consist of the same type of hexagonal layers, but the way the layers are stacked and connected is different. This might cause a tiny difference in the transversal relaxation during deformation.
 \begin{figure}[htbp]
 	\centering
 		\includegraphics[width=.6\textwidth]{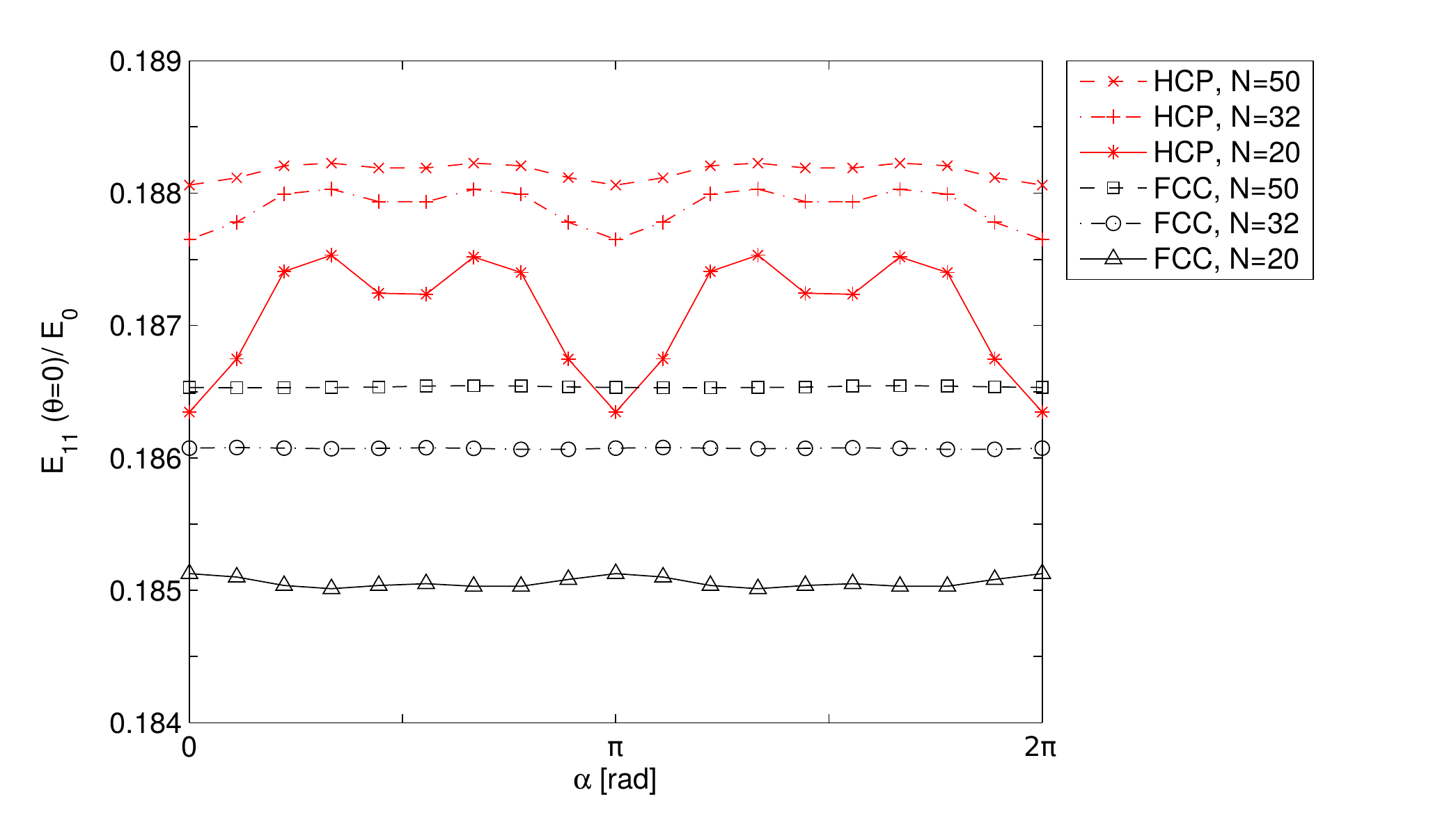} 
 	\caption{Demonstration of the numerical uncertainty of the Young's modulus in the \hkl (0 0 0 1) and \hkl (1 1 1) plane for HCP and FCCh, respectively, at $\phi_v = 0.635$. With increasing numerical resolution $N$ the Young's modulus becomes more isotropic. A small difference of 1 \% remains even for very high resolution. Note the small range of the vertical axis.}
 	\label{fig:hcp_fccH_iso}
 \end{figure}
\\The extremal values of the Poisson ratio and its dependence on the angle $\alpha$ are shown in Figure~\ref{fig:nu_max_min} and~\ref{fig:nu_winkel_kombi}, respectively, for a void fraction of $\phi_v = 0.635$. The \hkl (0 0 0 1) and \hkl (1 1 1) planes are isotropic in terms of the Poisson ratio. But there is a difference in the values between both structures, supporting the idea of a difference in the Young's modulus between the \hkl (0 0 0 1) and \hkl (1 1 1) plane. For FCCh the Poisson ratio $\nu_{21}$ is higher than for HCP. This means, that under strain orthogonal to the \hkl [1 1 1] direction FCCh relaxes more along the \hkl [1 1 1] direction than HCP along the \hkl [0 0 0 1] direction. Consequently, FCCh is softer in the \hkl (1 1 1) plane than HCP in the \hkl (0 0 0 1) plane. Still, both planes are isotropic in terms of the Poisson ratio.
We have recently published a more detailed study on the relations of the elastic parameters of HCP and FCC void material and their scaling behaviour~\cite{Heitkam2016}.

\afterpage{\clearpage}
\subsection{Random ordering}
In case of random void arrangement the mean Young's modulus, averaged over 45 different, randomly generated RVEs is shown in Figure~\ref{fig:E_max_min} for different void fractions. The mean Young's modulus is relatively low, compared to periodic structures. The reason is that in random packing column-like structures do not exist. Maximum and minimum Young's modulus for a given void fraction are derived by calculating the extremal values of each RVE separately and averaging them over all 45 corresponding RVEs. Note, that the extremal values depend on the numbers of voids within an RVE. A higher number of voids causes an averaging effect, yielding less prominent extremal values. For an infinite number of voids random packing is isotropic. However, the extremal values give an estimation of the local variation of the mechanical properties in random packing. As an example, Figure~\ref{fig:E_winkel_kombi} shows the dependence of the Young's modulus on the direction at $\phi_v = 0.425$ for one arbitrarily selected RVE. Even though only 30 voids are considered, the variations are small, below $3 \%$, and no distinct anisotropy is evident.
The Poisson ratio of random void packing is shown in Figure~\ref{fig:nu_max_min}. Since random packing should be isotropic in the limit of an infinite sample, the definition of the orthonormal basis $\left\{ \mathbf{f}_1, \mathbf{f}_2, \mathbf{f}_3 \right\}$ is arbitrary. However, for comparison with the other structures the same algorithm is used. Again, the variation of $\nu_{ij,max}$ is a measure for the local anisotropy of random packing. It is below $3 \%$ in the present case.\\
Computation of overlapping voids was not possible for random packing with the present method, because one has to define the boundaries of the RVE with sufficient distance to the void surfaces to create a suitable FE mesh. With increasing void diameter, possible positions for these boundaries become less available.

\afterpage{\clearpage}
 \section{Conclusions}
 \label{sec:conclusions}
The research question at the start of this investigation was whether a particular arrangement of spherical voids in a regular lattice would yield more advantageous properties than others, so that the fabrication process should target such an arrangement. Having achieved a level of uncertainty of 0.2 \%, the present study provides very sound data for the linear elastic properties of a number of such void materials. These are reported and a detailed comparison is undertaken. The result is that for given solid fraction, the variation of direction-averaged values among the structures is small. In this situation one may seek to extract useful structure-independent rules. For example, this study gives a fitting curve for the mean Young's modulus, averaged over all directions, that represents quite well a wide range of void fractions and different void arrangements.\\
The present study is confined to linear elasticity, but extension to non-linear behaviour of the interstitial material will be interesting to investigate. Non-linear behaviour is expected to be much more sensitive to the void organisation, resulting, e.g., from buckling phenomena~\cite{Gibson1997}. Another important aspect is the growth and coalescence of voids, leading to material failure~\cite{Tvergaard1990} which might also be sensitive to the void organization.\\
The dominant structural feature of these void materials consists of the necks confined between three or four voids, which narrow to zero thickness as rigidity-loss is approached. These necks or columns have been used throughout this study to explain elastic behaviour of different void materials. In an associated study~\cite{Heitkam2016} a beam model is proposed which takes into account the topology of the network of necks in HCP and FCC void material and derives a scaling theory for the elastic behaviour close to rigidity-loss. The simple beam model shows very good agreement with the detailed FE simulation in the present paper. A notable feature of the present results is the occurrence of Poisson ratios that lie outside the well known bonds for isotropic materials. This is also reproduced and well explained with the beam model.\\
%
%The structures considered here are motivated by theoretical considerations. Nevertheless they could also be fabricated. The related effort depends on the void fraction considered. High void fractions could be generated by liquid foam, possibly imposing additional \\
%
The present study is motivated by the elastic properties of solid foams. In case of monodisperse bubbles shaped boundaries~\cite{Gabbrielli2012} or electromagnetic fields~\cite{Heitkam2014} could create regularly arranged bubble crystals. However, the investigated geometries differ substantially from solid foam. Close to a gas fraction corresponding to touching bubbles the bubbles would remain approximately spherical, justifying the approach of spherical voids. With increasing gas fraction, bubbles in a foam deform. Seeking a state of minimum surface energy they form thin, flat lamellas between neighbouring bubbles and elongated Plateau borders confined between three bubbles. These Plateau borders correspond to the necks or columns, mentioned above, but show a substantially different geometry~\cite{Cantat2013}. This presumably yields a different elastic behaviour which is less sensitive to the solid fraction. Gibson and Ashby~\cite{Gibson1997} modelled an open-cell foam structure to consist of a network of straight beams with no necks but constant beam thickness. They found the Young's modulus to scale with the square of the void fraction while necks give a scaling to the power of 3.5, as shown in our other study~\cite{Heitkam2016}. The realistic scaling presumably lies somewhere between these extreme approximations.\\
For computations of more realistic foams at very high gas fractions, one could use the Surface Evolver~\cite{Brakke1992} to find realistic geometries. Subsequently, one may apply FE simulations similar to the present work, in order to extract the elastic behaviour of these foams. However, it is expected that the qualitative behaviour is similar. For these extensions the present study provides valuable reference data which clarify the physical properties and may also serve as validation data for further assessments.

%as well as the relations between different bubble arrangements should reflect the present study. To achieve the elastic behaviour computed in this study one needs to create void material with regularly arranged spherical voids. This is possible but very demanding with real foam. One needs to create bubbles of equal size. And one needs to enhance regular ordering of these bubbles in the desired arrangement. To enhance ordering one could apply shaped boundaries~\cite{Gabbrielli2012} or electromagnetic fields~\cite{Heitkam2014}. Without sufficient manipulation bubbles at low void fraction will adapt close-packed arrangements such as HCP and FCC. With decreasing liquid fraction, they will switch to BCC arrangement (Kelvin structure) because this structure corresponds to smaller surface energy. A structure with even lower surface energy would be the Weaire-Phelan structure~\cite{Weaire1994}, but this can only be achieved with sufficient templating~\cite{Gabbrielli2012}.
%Ordered void structure might be achieved using sacrificial templating. Spherical templates of equal size can be arranged in a crystalline lattice. The resulting void material would be limited to void fractions close to $\phi_{v,touch}$.  A promising approach to material with regularly arranged spherical voids would be 3D-printing, which even allows for further optimisation of the structure. To sum up, there are manufacturing routes to benefit from the special elastic properties of material with ordered spherical voids, but they are not straightforward. \\
%

%

 \afterpage{\clearpage}
 \section*{Acknowledgements}
We gratefully acknowledge fruitful discussions with Christophe Poulard. Computation time was provided by the Center for Information Services and High Performance Computing (ZIH) at TU Dresden. We acknowledge support from the European Research Council (ERC) under the European Union’s Seventh Framework Program (FP7/2007-2013) in form of an ERC Starting Grant, agreement 307280-POMCAPS. We acknowledge support from the European Centre for Emerging Materials and Processes (ECEMP) at TU Dresden and the Helmholtz-Alliance Liquid Metal Technologies (LIMTECH). DW acknowledges the support of SFI.


%\bibliographystyle{mhd}
%\bibliography{FEM_literatur}

\end{document}